\pdfoutput=1

\documentclass[reprint, amsmath, amssymb, aps, superscriptaddress]{revtex4-1}

\usepackage{graphicx}
\usepackage{dcolumn}
\usepackage{bm}
\usepackage{hyperref}
\usepackage{color}
\usepackage{lmodern}
\usepackage{epstopdf}

\begin{document}

\title{Sudden adiabaticity entering field-induced state in UTe$_2$}

\author{Rico Sch{\"o}nemann}
    \email{rschoenemann@lanl.gov}
    \affiliation{MPA-MAGLAB, Los Alamos National Laboratory, Los Alamos, New Mexico 87545, USA.}
\author{Priscila F. S. Rosa}
    \affiliation{MPA-Q, Los Alamos National Laboratory, Los Alamos, New Mexico 87545, USA.}
\author{Sean M. Thomas}
    \affiliation{MPA-Q, Los Alamos National Laboratory, Los Alamos, New Mexico 87545, USA.}
\author{You Lai}
    \affiliation{MPA-MAGLAB, Los Alamos National Laboratory, Los Alamos, New Mexico 87545, USA.}
\author{Doan N. Nguyen}
    \affiliation{MPA-MAGLAB, Los Alamos National Laboratory, Los Alamos, New Mexico 87545, USA.}
\author{John Singleton}
    \affiliation{MPA-MAGLAB, Los Alamos National Laboratory, Los Alamos, New Mexico 87545, USA.}
\author{Eric L. Brosha}
    \affiliation{MPA-11, Los Alamos National Laboratory, Los Alamos, New Mexico 87545, USA}
\author{Ross D. McDonald}
    \affiliation{MPA-MAGLAB, Los Alamos National Laboratory, Los Alamos, New Mexico 87545, USA.}
\author{Vivien Zapf}
    \affiliation{MPA-MAGLAB, Los Alamos National Laboratory, Los Alamos, New Mexico 87545, USA.}
\author{Boris Maiorov}
    \affiliation{MPA-MAGLAB, Los Alamos National Laboratory, Los Alamos, New Mexico 87545, USA.}
\author{Marcelo Jaime}
    \email{mjaime@lanl.gov}
    \affiliation{MPA-MAGLAB, Los Alamos National Laboratory, Los Alamos, New Mexico 87545, USA.}
 
\date{\today}

\begin{abstract}
There has been a recent surge of interest in UTe$_2$ due to its unconventional magnetic field (H) reinforced spin-triplet superconducting phases persisting at fields far above the simple Pauli limit for H $\parallel$ [010]. Magnetic fields in excess of 35 T then induce a field-polarized magnetic state via a first-order-like phase transition. More controversially, for field orientations close to H $\parallel$ [011] and above 40 T, electrical resistivity measurements suggest that a further superconducting state may exist. However, no Meissner effect or thermodynamic evidence exists to date for this phase making it difficult to exclude a simple low-resistance metallic state. In this paper, we describe a study using thermal, electrical, and magnetic probes in magnetic fields of up to 55 T applied between the [010] ($b$) and [001] ($c$) directions. Our MHz conductivity data reveal the field-induced state of low or vanishing electrical resistance; simultaneous magnetocaloric effect measurements (i.e. changes in sample temperature due to changing magnetic field), show the first definitive evidence for adiabaticity and thermal behavior characteristic of bulk field-induced superconductivity.  
\end{abstract}

\maketitle

\section{Introduction}

The recently discovered actinide superconductor UTe$_{2}$ has been predicted as a promising candidate for the realization of chiral spin-triplet superconductivity with equal-spin pairing. Support for this picture comes from its close proximity to magnetic order, its unusually large critical magnetic field (far exceeding the Pauli limit for a weakly coupled BCS superconductor in the absence of spin-orbit coupling), as well as the observation of only a small change in the Knight shift below its superconducting transition temperature $T_{\mathrm{c}} \approx 1.6-2.1$~K~\cite{ran_nearly_2019, ran_extreme_2019, knafo_magnetic-field-induced_2019, knebel_field-reentrant_2019, nakamine_superconducting_2019, jiao_stm_2020, aoki_unconventional_2022, rosuel_2023, matsumura_2023}. UTe$_{2}$ crystallizes in a body-centered orthorhombic structure ($Immm$)~\cite{ran_nearly_2019}. Unlike closely related orthorhombic {UGe$_{2}$, URhGe and UCoGe, for which superconductivity emerges within the ferromagnetically ordered state \cite{aoki_unconventional_2022}, no signs of superconductivity coexisting with magnetic order were observed in UTe$_{2}$ down to $25$~mK~\cite{sundar_coexistence_2019, paulsen_anomalous_2021}. Magnetic fluctuations are believed to play a major role in facilitating superconductivity in UTe$_{2}$ \cite{aoki_unconventional_2022}, yet the nature of the fluctuations is still a matter of contention. Indeed, whilst some experiments give evidence for ferromagnetic fluctuations~\cite{ran_nearly_2019}, recent neutron scattering data show excitations at an antiferromagnetic wave-vector~\cite{duan_incommensurate_2020}. Studies under hydrostatic pressure also support the presence of antiferromagnetic fluctuations \cite{thomas_evidence_2020, aoki_multiple_2021}. 

When a magnetic field is applied along the magnetically hard $b$-axis, a reinforcement of superconductivity is observed above $15$~T, which is extended up to $\mu_0 H_{\mathrm{m}} \approx 35$~T\cite{rosuel_2023} establishing the reinforcement of the critical field. At 35$~T$, a first-order metamagnetic transition into a field-polarized paramagnetic phase occurs below 8~K, leading to a jump of $0.5\,\mu_{\mathrm{B}}$ in the magnetization and the termination of the superconducting state~\cite{ran_extreme_2019, knebel_field-reentrant_2019, miyake_metamagnetic_2019}. A smaller anomaly around $6.5\,\mathrm{T}$ was also reported in magnetization data for $H\parallel [100]$~\cite{miyake_metamagnetic_2019}. 
It is likely that a Fermi-surface reconstruction, as well as a volume/valence change, accompanies the metamagnetic transition~\cite{miyake_magnetovolume_2022}. Thermopower and Hall data show a change of the majority charge and heat carriers from electrons to holes with a step-like increase in the electrical resistivity~\cite{knafo_magnetic-field-induced_2019}. Based on the Hall data, the estimated carrier density for $H> H_{\mathrm{m}}$ is around a factor of six lower than that for $H < H_{\rm m}$~\cite{niu_evidence_2020}. 

On rotating the magnetic field from $H \parallel$ [010] towards $H \parallel$ [001], the metamagnetic transition at  $H_{\mathrm{m}}$ shifts upwards in field. Interestingly, when the field lies in a narrow angular range around the [011] direction {\it i.e.,} $\approx 23.7^{\circ}$ away from the [010] axis), transport measurements suggest that a state with an undetectable low resistance emerges within the field polarized paramagnetic phase above $H_{\mathrm{m}}$. This state has been interpreted as superconductivity \cite{ran_extreme_2019, knafo_comparison_2021, aoki_unconventional_2022, helm_arxiv_2023}, see Fig.[\ref{fig:figure1}(a)]; however, it is impossible to tell from simple transport experiments if the high-field, low-resistance, state is filamentary or bulk in nature. Constructing a theoretical model for superconductivity in strongly correlated electron systems has proven to be a phenomenally complex task. Until its bulk nature is established, extrinsic effects such as local stoichiometry, strain, lattice defects, impurities, etc. cannot be ruled out; all bets to explain the phenomenon are off. Crucial bulk thermodynamic evidence is still rather scarce though. Indeed, thus far low-temperature thermodynamic measurements have focused on magnetic fields along the principal axis $a$, $b$, and $c$, yet none for $H \parallel$ [011]  \cite{imajo_thermodynamic_2019, rosuel_2023}.

Here we report complementary proximity detector oscillator (PDO)~\cite{altarawneh_proximity_2009, ghannadzadeh_measurement_2011}, magnetocaloric effect (MCE)~\cite{jaime_high_2002, silhanek_irreversible_2006}, and angular dependent torque magnetometry measurements to $55\,\mathrm{T}$ that are eminently applicable in this context, as they provide together an unambiguous thermodynamic detection of phase transitions and were conducted in the pulsed magnetic fields required to observe the high-field phase. Our combined results show the first bulk evidence for a state characterized by extremely high electrical conductivity, reversible increase in temperature, and thermal decoupling from the bath, likely due to fully gaped bulk superconductivity.

\begin{figure}
\includegraphics{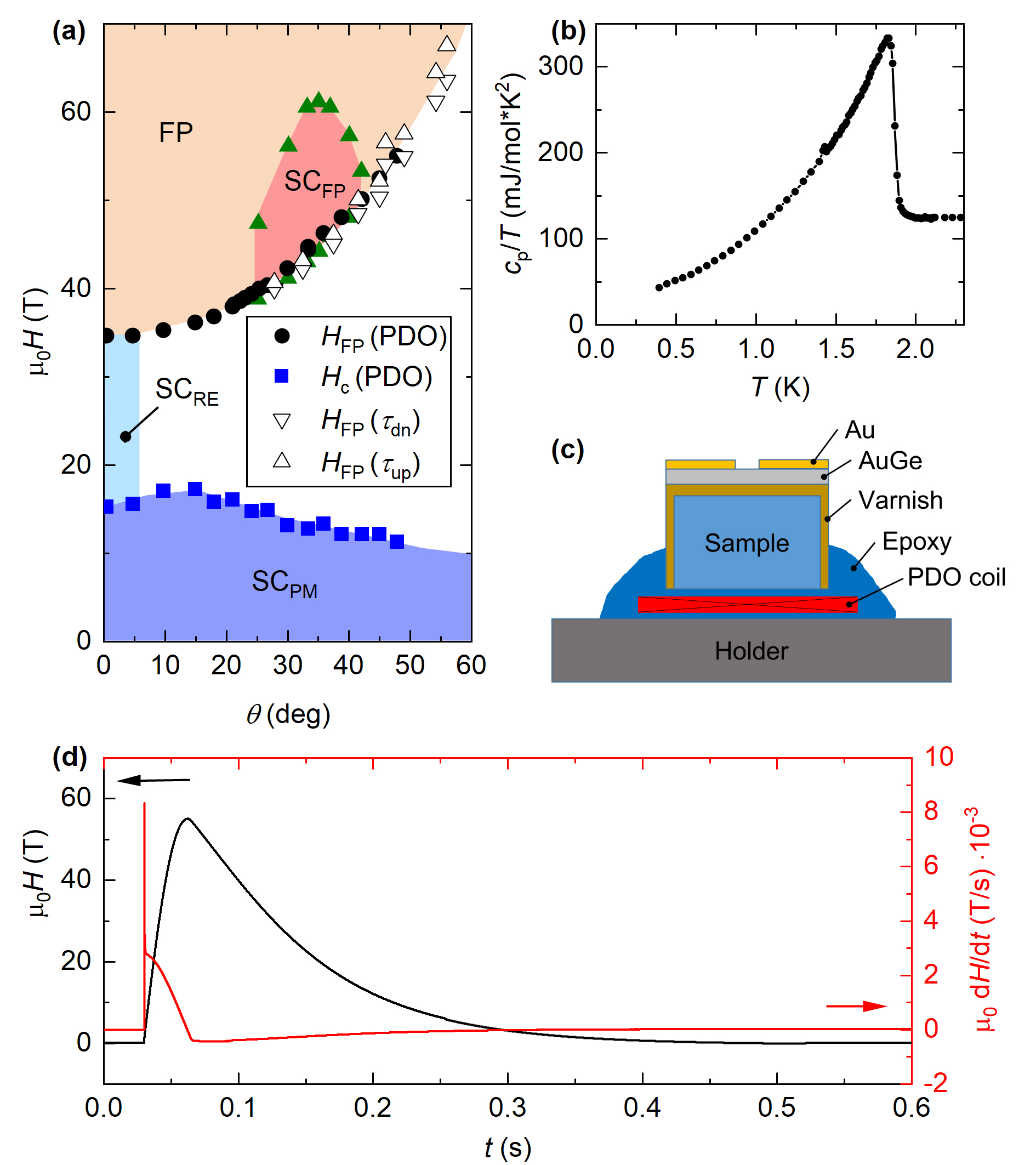}
\caption{(a) Low temperature field-angle ($H-\theta$) phase diagram of UTe$_{2}$, where $\theta = 0^{\circ}$ corresponds to $H\parallel b$ and $\theta = 90^{\circ}$ to $H\parallel c$. Blue squares mark the transition from the superconducting ground state ($\mathrm{SC_{PM}}$) to the paramagnetic state or the re-entrant superconducting state ($\mathrm{SC_{RE}}$) measured at $T \approx 0.6\,\mathrm{K}$. Note that the region of the $\mathrm{SC_{RE}}$ state is just roughly indicated to extend to $\theta \simeq 5^{\circ}$. The black circles and open triangles denote the first-order transition into the field-polarized paramagnetic state (FP). Critical fields were obtained from PDO (black circles, $T\approx 0.9\,\mathrm{K} $) and torque $\tau_{\mathrm{up/dn}}$ measurements (black open triangles, for up- and down-sweep, $T\approx 0.7\,\mathrm{K}$) in this work. The green triangles encircle the proposed high-field superconducting phase ($\mathrm{SC_{FP}}$); points were taken from Ran \textit{et al.} \cite{ran_extreme_2019}. (b) Specific heat vs. temperature for a UTe$_{2}$ single-crystal synthesized in the same batch as the ones used for MCE measurements. (c) Schematic of the sample arrangements for simultaneous MCE and PDO measurements in pulsed fields. (d) Magnetic field $H$ (black line) and $\mathrm{d}H/\mathrm{d}t$ (red line) as a function of time for the pulsed magnet used in the MCE measurements.}
\label{fig:figure1}
\end{figure}

\section*{Results}
Fig.~\ref{fig:figure1}(a) shows the phase diagram of UTe$_{2}$ for magnetic field magnitude and orientation ($H-\theta$); with the angle $\theta$ describing the field rotating from parallel to the crystallographic $b$-axis ($\theta = 0^{\circ}$) towards the $c$-axis ($\theta = 90^{\circ}$). The phase diagram is based on prior magnetization, electric and thermal transport measurements~\cite{ran_nearly_2019, ran_extreme_2019, knafo_magnetic-field-induced_2019, knebel_field-reentrant_2019, knafo_comparison_2021, aoki_unconventional_2022}; the points surrounding the high-field phase $\mathrm{SC_{FP}}$ were taken from Ran \textit{et al.}~\cite{ran_extreme_2019}. Despite sample temperature $T$ excursions of up to $\approx 1.0$~K (described in detail below), far from equilibrium with the $^3$He bath temperature ($\approx 0.6$~K), the field positions of both the high-field metamagnetic and low-field transition out of the SC$_{\rm PM}$ phase obtained from our PDO and MCE measurements are in good agreement with prior data. In the case of the metamagnetic transition, this is unsurprising as $H_{\rm m}$ is virtually temperature independent for $T<4$~K~\cite{aoki_unconventional_2022}. 

\begin{figure}
\includegraphics{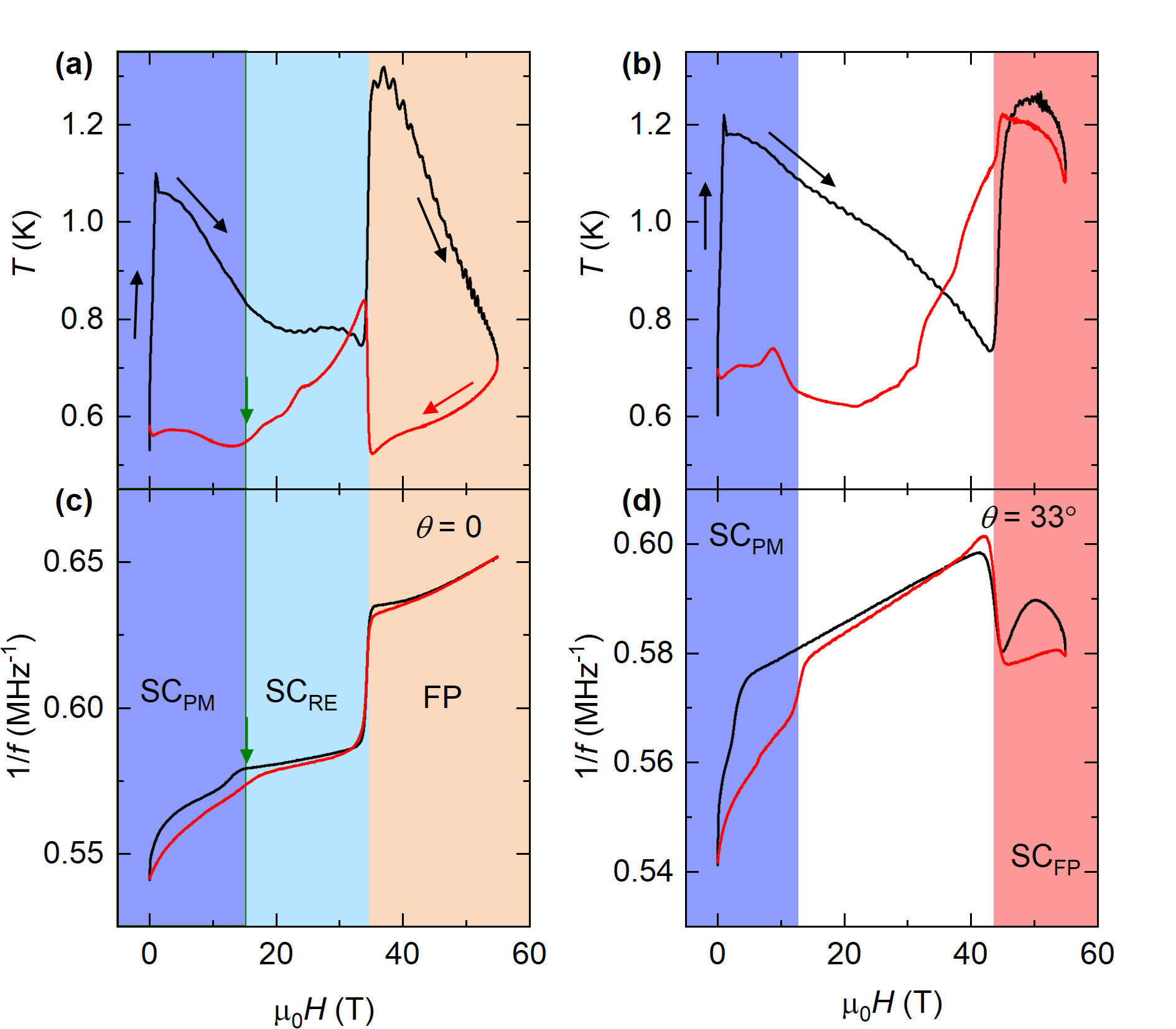}
\caption{(a) Sample temperature $T$ versus field data for $H\parallel b$ ($\theta = 0$). Data for the up-sweep (rising field) portion of the field pulse are shown in black and those for the down-sweep (falling field) are in red. The background color indicates the superconducting and magnetic phases displayed in Fig. \ref{fig:figure1}(a). Note that the critical fields for the low-field superconducting state (arrows) are marked for the down-sweep curves. (b) $T$ versus $H$ data for $\theta=33^\circ$. (c, d) PDO data, shown as inverse frequency $1/f$, a proxy for electrical resistance, versus applied field, was recorded simultaneously with the thermal measurements shown in (a, b). The color scheme is the same as in (a, b).}
\label{fig:figure2}
\end{figure}

Examples of sample temperature $T$ versus field curves for $H\parallel b$ ($\theta = 0$) and $\theta = 33^{\circ}$ are shown in Fig.~\ref{fig:figure2}(a, b) on top of data from simultaneous PDO measurements (c, d). Referring to the phase diagram [Fig.~\ref{fig:figure1}(a)], at sub-Kelvin temperatures and $\theta =0$, the up-sweep of a 55~T field pulse first traverses the low-field SC$_{\rm PM}$ phase, then the so-called re-entrant superconducting phase (SC$_{\rm RE}$) and the metamagnetic transition at $H_{\rm m}$ before finally entering the field-polarized (FP) (non-superconducting) phase. By contrast, at $\theta =33^\circ$, a similar pulse goes through the SC$_{\rm PM}$ phase, a metallic (paramagnetic, non-superconducting) phase, and the metamagnetic transition (shifted to higher fields), where it enters the SC$_{\rm FP}$ phase. As we see, these different paths across the phase diagram result in different thermal responses.

Turning first to the MCE data at $\theta = 0$ [Fig.~\ref{fig:figure2}(a)], as $H$ initially rises (black curve) there is a steep increase in $T$ from the $^3$He bath temperature ($\approx 0.6$~K) to $\approx 1.1$~K. This heating is attributable to an avalanche-like, dissipative vortex movement in the superconducting SC$_{\rm PM}$ phase, a phenomenon frequently seen in pulsed-field measurements of more conventional superconductors ({\it e.g.}, Ref.~\cite{SMYLIE}). Thereafter, $T$ relaxes towards the bath temperature until a sharp step upwards denotes the first-order phase transition at $H_{\rm m}$. Once in the FP state, $T$ again relaxes for the rest of the up-sweep and during the start of the down-sweep (red curve). However, at $H_{\rm m}$ on the down-sweep there is another sharp increase in $T$, followed by further relaxation down to around 15~T; below $\approx 13\,\mathrm{T}$ there is a gentle increase in $T$, likely due to a combination of SC gap opening and dissipative vortex motion as the removed field enables the SC$_{\rm PM}$ phase. Note that the down-sweep of $H$ is much slower than the up-sweep, allowing more time for heat generated to dissipate~\cite{SMYLIE}. The most significant results for $\theta =0^\circ$ are (a) the irreversible processes that cause heating at the metamagnetic transition regardless of field-change direction, which dominates the thermodynamics (likely due to this phase boundary being mostly temperature-independent \cite{aoki_unconventional_2022, silhanek_irreversible_2006}; (b) the thermalization (cooling off) of the sample in the high- field/high-resistance FP state. Here the field changes do not cause eddy-current heating, and thermal coupling to the bath dominates the sample thermal response, and (c) irreversible and reversible processes associated with the low field SC$_{\rm PM}$ phase. 

The simultaneous PDO data at $\theta =0$ [Fig.~\ref{fig:figure2}(c)] reflect these $T$ changes. As the field increases (black curve) there is a sharp fall in $f$ at about 15~T, indicating the SC$_{\rm PM}$ to SC$_{\rm RE}$ transition (the corresponding $T$ versus $H$ curve in (a) flattens at about the same field). The sample exits the SC$_{\rm RE}$ phase at $\mu_{0} H_{\mathrm{m}}=35\,\mathrm{T}$; once in the non-superconducting FP phase, shifts $\Delta f$ in the PDO frequency are dominated by changes in the sample resistivity $\Delta \rho$, with an approximate proportionality $\Delta f \propto 1/\Delta \rho$~\cite{ran_extreme_2019, altarawneh_proximity_2009,ghannadzadeh_measurement_2011}. At $H_{\rm m}$, $\rho$ is known to exhibit a sharp increase~\cite{ran_extreme_2019}, leading to a downward step in $f$\footnote{In the PDO circuit, the resonant frequency $f = 1/\sqrt{LC}$, where $L$ is the circuit inductance and $C$ the capacitance. The inductance of a solenoid of $N$ turns, area $A$, and length $l$ is $L=\mu N^2A/l$, where $\mu$ is the permeability of the core material. Due to the skin-depth effect in metallic samples of low resistance the effective volume, and hence inductance, of the coil is reduced. Hence, $\Delta f \propto (\Delta\rho)^{-1}$. In UTe$_2$ this frequency shift is most evident upon entering the superconducting state where the penetration depth is significantly smaller than the normal state skin depth.}. Above $H_{\rm m}$, the normal-state resistivity of UTe$_2$ is rather $T-$independent in the range $0.6-2$~K~\cite{ran_extreme_2019}; hence, despite the varying $T$ seen in the MCE data, the PDO frequency on the downsweep of the field (red curve) overlies the up-sweep data. Below $H_{\rm m}$, slight hysteresis between down-sweep (red) and up-sweep PDO data marks the presence of the SC$_{\rm RE}$ phase before a step upwards (marked by an arrow) shows the transition back to the SC$_{\rm PM}$ phase; as $T$ is lower on the down-sweep [Fig.~\ref{fig:figure2}(a)], this latter feature occurs at a slightly higher field than the corresponding feature in the up-sweep.

\begin{figure}
\includegraphics{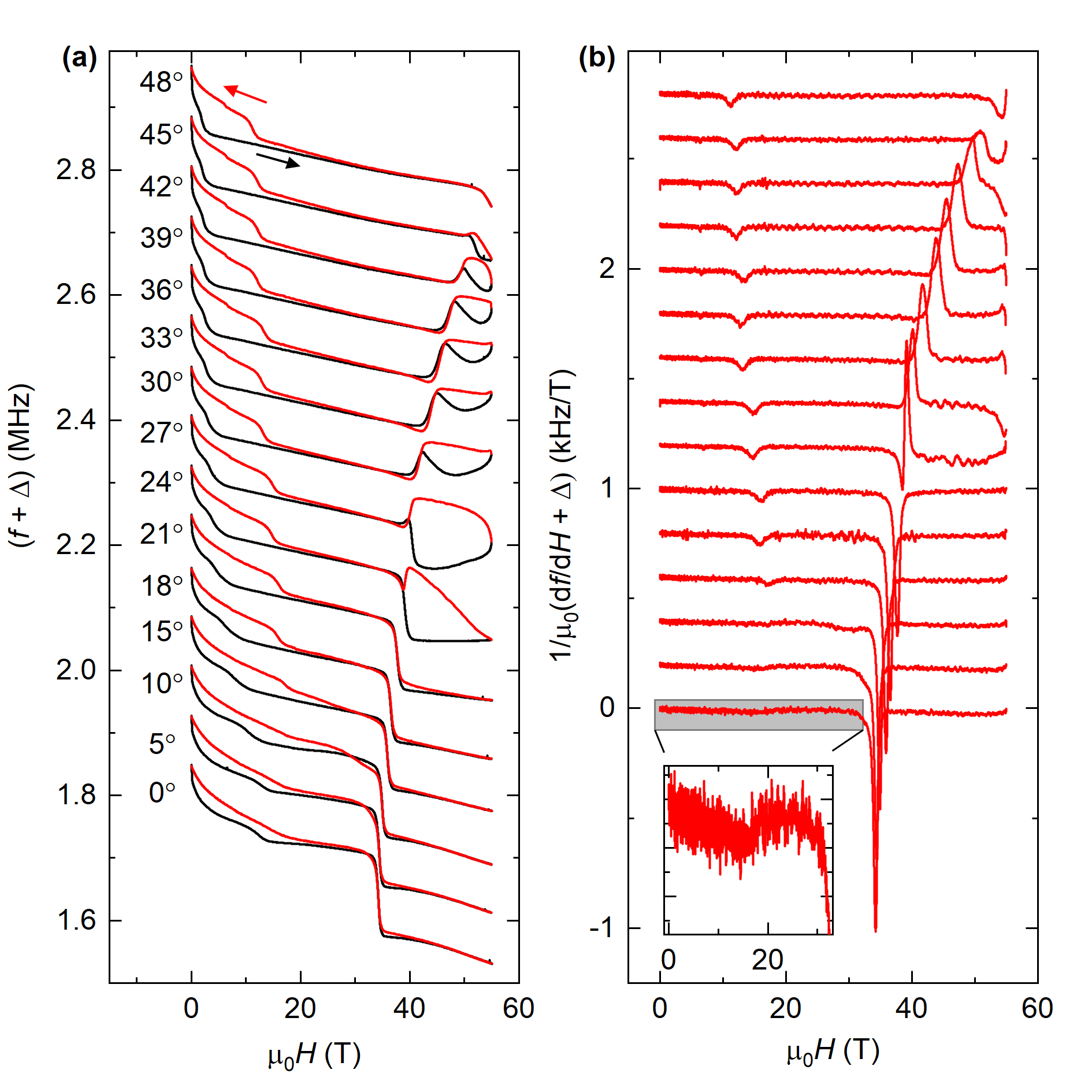}
\caption{(a) PDO frequency (a proxy for electrical conductivity) vs. magnetic field for different angles $\theta$ displayed on the left of each curve. The field up- and down-sweeps are shown as black and red curves respectively. (b) Derivative of the down-sweep curves shown in (a). The inset shows a low-field feature indicating the transition between the SC$_{\rm PM}$ and SC$_{\rm RE}$ superconducting states close to $\theta = 0$. Curves in (a) and (b) are shifted vertically by an offset $\Delta$ for clarity.}
\label{fig:figure3}
\end{figure}

\begin{figure*}
\includegraphics{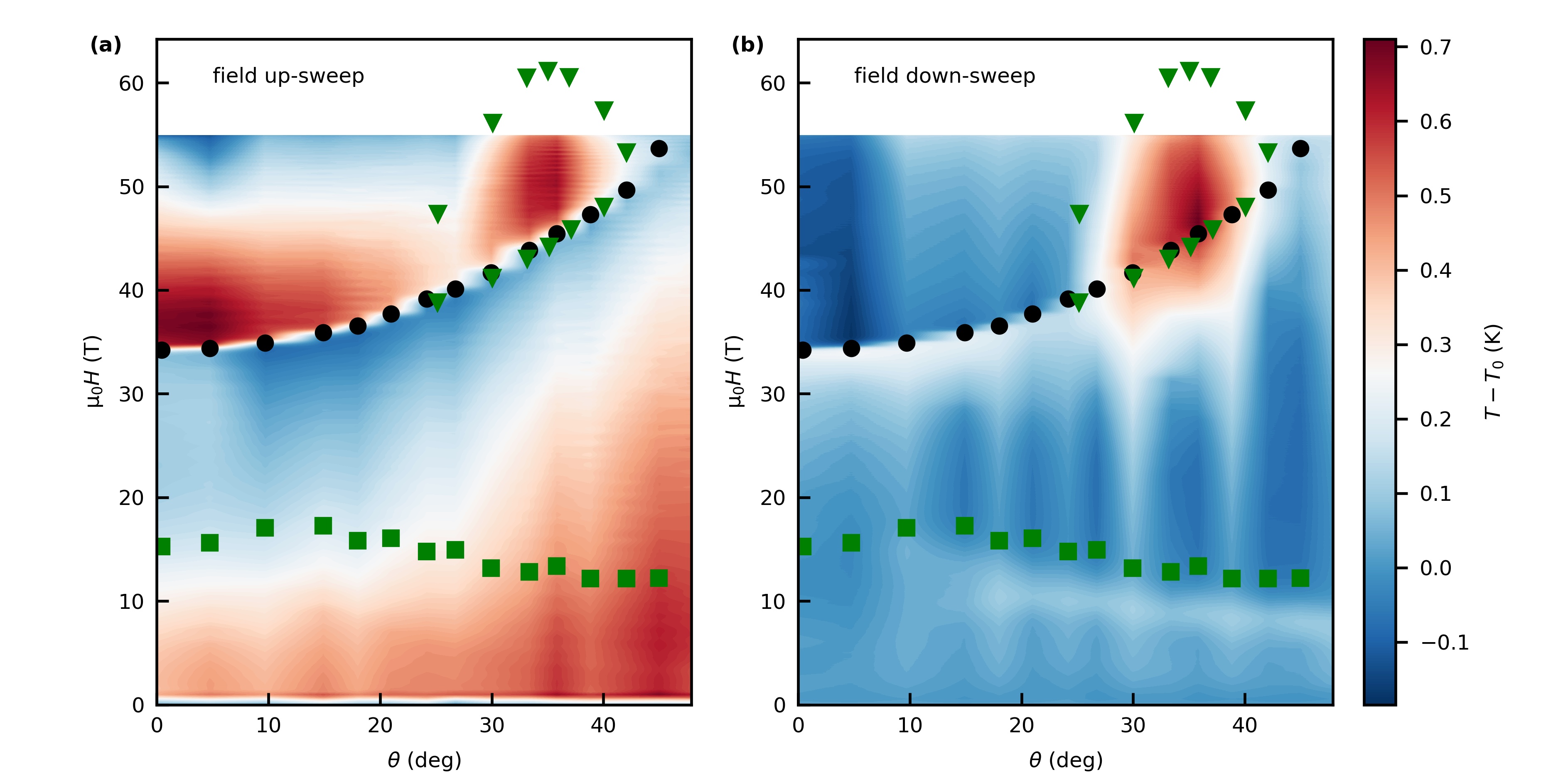}
\caption{Contour plot of the sample temperature $T$ as a function of the angle $\theta$ at which the magnetic field $H$ is applied for the up-sweep (a) and down-sweep (b). The black circles indicate the metamagnetic phase transitions discussed in the text, green squares and triangles enveloping the high-field superconducting state were taken from Ran \textit{et al.}~\cite{ran_extreme_2019}. The reversible temperature increase and adiabaticity observed for H $>$ 40 T and 30$^o < \theta < 40^o$, quite distinct from results at all other angles and fields, coupled with high electrical conductivity, are together consistent with a field-induced superconducting state in UTe$_2$. The initial temperature before the field pulses is $T_{0} \sim (0.6\pm 0.1)\,\mathrm{K}$, variations in $T_{0}$ cause vertical stripes to appear in both contour plots.}
\label{fig:figure4}
\end{figure*}

The MCE and PDO data for $\theta = 33^{\circ}$ [Fig.~\ref{fig:figure2}(b,d)], below about 15~T, behave in a similar way to their counterparts at $\theta = 0$. However, the lack of the SC$_{\rm RE}$ phase at $\theta =33^\circ$ means that $T$, rather than flattening, continues to fall until $H_{\rm m}$ is reached. Correspondingly, the PDO signal above 15~T at $\theta = 33^\circ$ decreases roughly linearly, reflecting the increasing normal-state magnetoresistance, rather than flattening out as it did at $\theta = 0$ due to the presence of the SC$_{\rm RE}$ phase. However, the biggest contrast for $\theta = 33^\circ$ compared to $\theta = 0$ occurs on crossing $H_{\rm m}$. A temperature increase slightly less abrupt is seen followed by a temperature drop inside the SC$_{\rm FP}$ phase which, quite remarkably, retraces itself during the field down sweep in an adiabatic fashion. The cooling continues upon crossing the phase boundary back to the paramagnetic normal state. (The full angular dependence of the MCE at $H_{\rm m}$ is discussed further along). Continuing along the down-sweep curves, the $33^\circ$ PDO data show an increase in $f$ due to the normal-to-SC$_{\rm PM}$ transition, accompanied by slight heating due to vortex motion revealed by the MCE data. The most significant results for $\theta = 33^{\circ}$ are  (a) The largely reversible change in temperature observed at $H_{\rm m}$ with minimal dissipative mechanisms. Here a temperature increase on the upsweep is suggestive of the opening of an energy gap for excitations. The drop in temperature during the downsweep marks concomitantly the reversible closing of the gap. (b) The re-tracing of the sample temperature inside the pink shadow region (upsweep and downsweep overlap) indicates adiabaticity, also compatible with a gaped state where superconducting pairs carry no entropy, decoupling the sample from the thermal bath. Note that a hypothetical low-resistance metallic state would result in enhanced eddy-current heating for both upsweep and downsweep traces (lifting the overlap) in the high field state, a result very different from our observations. 

Having described the signatures of the various phase boundaries in the PDO and MCE data, we now turn to 
Fig.~\ref{fig:figure3}(a), which shows PDO frequencies for 15 angles in the range $0\leq \theta \leq 48^\circ$; as before, black curves signify rising $H$ and red curves falling $H$. Note that the field at which the drop in $f$ associated with the exit from the SC$_{\rm PM}$ phase (either into the SC$_{\rm RE}$ phase $(\theta \leq 10^\circ)$ or normal state $(\theta > 10^\circ)$ occurs at lower fields on the field up-sweep due to the heating seen in the MCE experiment; the sample is much closer to the bath temperature on the downsweep, so that the corresponding step is at higher fields~\cite{SMYLIE}. 

Corresponding derivatives $(1/\mu_0)(\mathrm{d} f/\mathrm{d}H)$ of the down-sweep data are shown in Fig.~\ref{fig:figure3}(b). The critical fields $H_{\mathrm{c}}$ and $H_{\mathrm{FP}}$ shown in Fig. \ref{fig:figure1}(a) were extracted from the extrema of this data set.  For the three lowest $\theta$ values $(0, 5^\circ, 10^\circ)$ there is only a weak, broad feature between 15 and 20~T, reflecting that the transition is between two superconducting phases (SC$_{\rm PM}$ and SC$_{\rm RE}$). For $\theta> 10^\circ$, the weak feature is replaced by a well-defined minimum, as it now corresponds to a superconductor (SC$_{\rm PM}$)-to-normal transition.  

The MCE measurements are summarized in Fig.~\ref{fig:figure4}; the increase in $T$ in the SC$_{\rm FP}$ phase around $\theta = 33^{\circ}$ clearly stands out, staying hot in the down-sweep data. This provides thermodynamic evidence that the sample becomes thermally decoupled from the bath (indeed, the SC$_{\rm FP}$ region does not change color for upsweep and downsweep, unlike the rest of the H,$\theta$ phase space), and a compelling proof for the bulk nature of the SC$_{\rm FP}$ state observed in UTe$_{2}$. On the other hand, due to the large heating effect caused by vortex motion at the onset of the magnet pulse, near $H$=0,  no clear phase boundary of the low field superconducting phase can be identified in the up-sweep MCE data. Based on the PDO data (Fig. \ref{fig:figure3}), the SC$_{\rm PM}$ phase is suppressed at a field of a few Tesla on the up-sweep. During the down-sweep, the phase boundary into the SC$_{\rm PM}$ phase coincides with the onset of gentle sample heating below $\approx 15\,\mathrm{T}$ and the corresponding upward step in the PDO data (Fig. \ref{fig:figure3}).

At the close of this section, we emphasize that though the corresponding features in the PDO and MCE data are weak, there are distinct indications of the boundary between the SC$_{\rm PM}$and SC$_{\rm RE}$ phases. This seems to confirm that though both states are superconducting, they are distinct phases with subtly different properties~\cite{aoki_unconventional_2022, rosuel_2023}.

\begin{figure}
\includegraphics{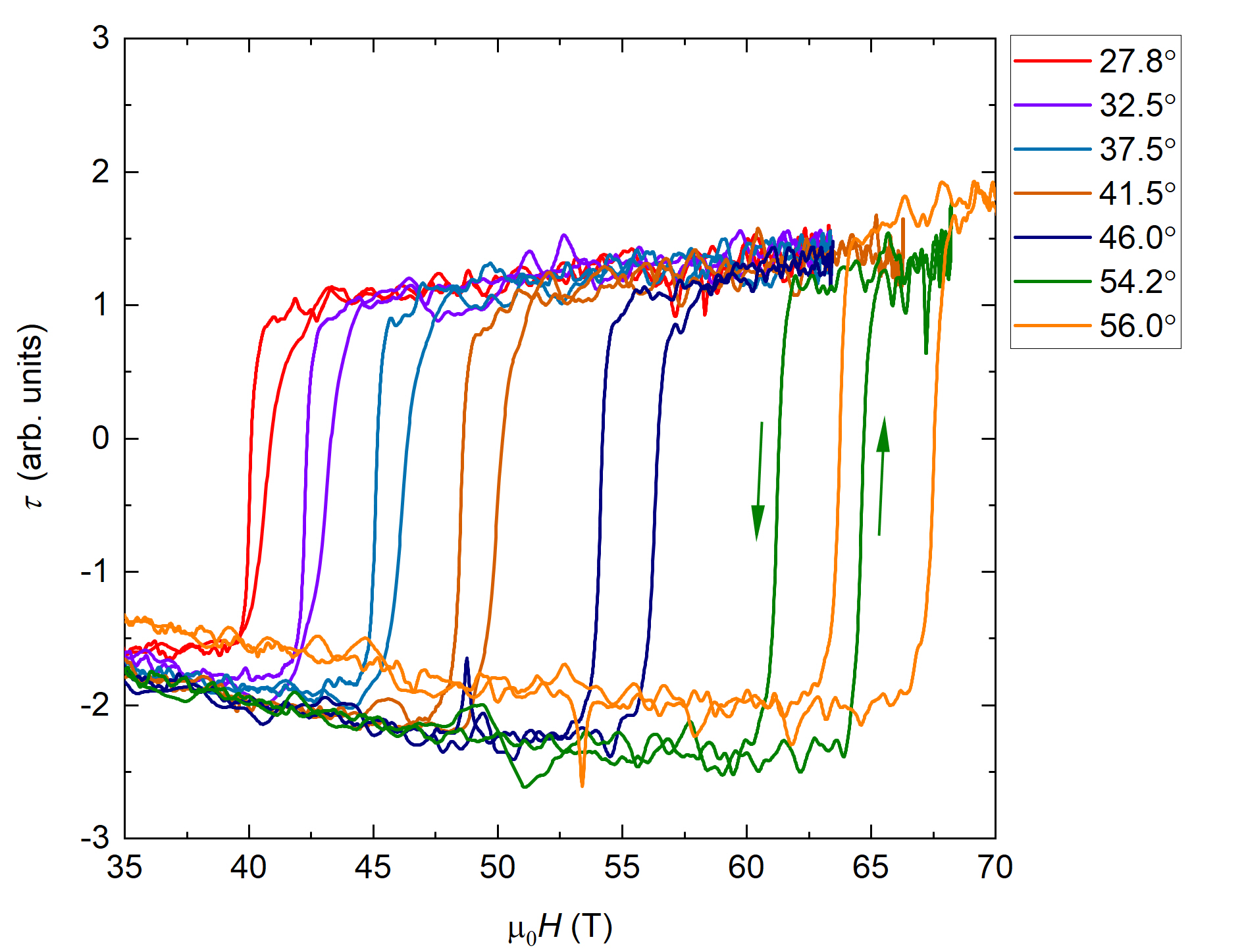}
\caption{Metamagnetic transition in UTe$_{2}$ as seen in the magnetic torque signal measured at different angles at $T=(0.7\pm 0.1)\,\mathrm{K}$. Up- and down-sweep of the magnetic field are indicated by arrows. In general, the curve with the higher transition field corresponds to the upsweep.}
\label{fig:figure6}
\end{figure}

\section*{Discussion}
Before treating the thermodynamics of the onset of the high-field SC$_{\rm FP}$ state in more detail, it is worth considering whether there is an alternative explanation for the previous (non-thermodynamic) data used to identify the apparent superconductivity of this phase. 

One possibility might be a low (but nonzero) resistivity metallic phase caused by a field-induced Fermi-surface reconstruction at $H_{\rm m}$ that occurs over a restricted range of field orientations. However, existing experimental data provide a number of objections to such an interpretation.
\begin{enumerate}
\item
As mentioned in the Introduction, Hall-effect and thermopower measurements~\cite{knafo_magnetic-field-induced_2019}for $H \parallel b$ indicate a very significant {\it decrease} in the charge-carrier density as one crosses $H_{\rm m}$ into the FP (normal) state, leading to a strong {\it increase} in the resistivity~\cite{knafo_magnetic-field-induced_2019,ran_extreme_2019}.
\item
To counter the previous point, one might argue that a significantly different change in electrical properties ({\it i.e.}, a large {\it increase} in carrier density and/or mobility) occurs at $H_{\rm m}$, but only over a special, restricted range of $\theta$. In such a case, one would expect that the metamagnetic transition would also change in character for these angles. However, torque magnetometry data (Fig. \ref{fig:figure6}) carried out over a wide range of field orientations show that the position and size of the magnetization jump at $H_{\rm m}$ vary smoothly and monotonically with $\theta$.
\item
An increase in the charge-carrier density at $H_{\rm m}$ (such as due to the closing of an energy gap) leads to cooling (see \cite{jaime_closing_2000, jaime_frontiers_2009} for Ce$_3$Bi$_4$Pt$_3$ and \cite{jaime_high_2002} for URu$_2$Si$_2$) of the sample during the field up-sweep and heating in the down-sweep, which is incompatible with the data in this report. A highly conductive field-induced gap-less metallic state would also likely lead to eddy-current heating in changing fields of both directions, which is not present in the MCE data.
\item 
The observation of reversible heating upon entering the high field SC$_{\rm FP}$ phase indicating the opening of an energy gap in excitations, is accompanied by sample thermal decoupling (adiabaticity). This is the consequence of an increase in the sample thermal relaxation time constant $\tau \propto C/K$, where $C$ is the sample heat capacity  and $K$ is its thermal conductance. Since the opening of an energy gap leads to a reduction of $C$, the observed increase in  $\tau$ points to a significant drop in $K$ reminiscent of a superconducting state, as seen in numerous U-based compounds such as UBe$_{13}$, UPt$_3$, UCoGe, and URhGe  \cite{ravex_1987, suderow_1997, howald_PhD, taupin_2014}, as well as in UTe$_2$ in low fields \cite{rosuel_2023}.    

\item
The PDO data used to detect the SC$_{\rm FP}$ state in Ref.~\cite{ran_extreme_2019} (and those in this paper) behave in a qualitatively similar manner to PDO measurements on more conventional superconductors such as pnictides~\cite{SMYLIE, NIKOLO} and cuprates~\cite{CdC}, especially in the hysteresis observed between up-sweeps and down-sweeps of the field. By contrast, PDO data measured in systems where there is a large field-induced increase in carrier density but no superconductivity~\cite{Xiang2021, PCHo} behave in a very different way, e.g. showing different hysteretic characteristics.
\item
The typical energy scales associated with the transition at $H_{\rm m}$ are $\sim 40~$K (see {\it Introduction} above).
Any phenomenon associated with increased (normal-state) conductivity  due to a Fermi-surface change at $H_{\rm m}$ would be expected to persist (or slowly die away) over a temperature range similar to this. By contrast, the upper-temperature limit of the SC$_{\rm FP}$ phase is about 1.9~K~\cite{ran_extreme_2019}, very similar to the critical temperatures of the SC$_{\rm PM}$ and SC$_{\rm RE}$ superconducting phases~\cite{aoki_unconventional_2022}, suggesting a common or closely related origin.
\end{enumerate}

Yet another proposal for the field-induced state is that of an sliding density wave (sDW), as observed in low-dimensional materials \cite{stokes_1984} when a depining electric field is applied. Since the Fermi surface in UTe$_2$ is quasi-2D, a sDW cannot be ruled out from transport experiments, in part because these are run in the presence of an electrical current and, hence, an electrical field on the sample. Our MCE experiments are, however, conducted in zero-current/zero-electrical field and a sDW condensate is unlikely.    

In view of the above points, the following discussion of the thermodynamics occurring at and around $H_{\mathrm{m}}$ assumes that the SC$_{\rm FP}$ phase is superconducting.

As shown in a previous study~\cite{imajo_thermodynamic_2019} for $H \parallel b$, the metamagnetic transition at $H_{\rm m}$ is first-order at low temperatures and accompanied by hysteresis losses. In the current, field-orientation-dependent study, the temperature change $\Delta T_{\mathrm{FP}}$ observed at $H_{\rm m}$ can be described as follows (see  Fig.~\ref{fig:figure5}). (i)~During the up-sweep, $\Delta T_{\mathrm{FP}}$ is positive and decreases with increasing $\theta$ (dashed line in Fig.~\ref{fig:figure5}(b)). (ii)~$\Delta T_{\mathrm{FP}}$ increases for $\theta$ between $25^{\circ}$ and $35^{\circ}$ as the sample transitions into the SC$_{\rm FP}$ state. (iii)~$\Delta T_{\mathrm{FP}}$ decreases with $\theta$ once again when the SC$_{\rm FP}$ state is suppressed at larger $\theta$. During the down-sweep of the field, $\Delta T_{\mathrm{FP}}$ (Fig.~\ref{fig:figure5}(b), red points) is always smaller than that during the up-sweep. For the falling field, $\Delta T_{\mathrm{FP}}$ is positive for $\theta < 27^{\circ}$ and becomes negative for larger angles. 

In making a quantitative description of the thermodynamics of the metamagnetic transition, we assume that the overall entropy change is a sum of reversible and irreversible processes,

\begin{equation}
    \Delta S = \Delta S_{\mathrm{rev}} + \Delta S_{\mathrm{irr}} = \frac{C_{p}\Delta T}{T} + \frac{\partial Q_{\mathrm{loss}}}{T}.
\end{equation}

\noindent where $\Delta S_{\mathrm{rev}}$ describes the latent heat released during the transition, which is recovered when the field crosses $H_{\mathrm{m}}$ in the opposite sense; and $C_P$ is the heat capacity at constant pressure. The small field width of the metamagnetic transition leads us to assume adiabatic conditions and extract the temperature change $\Delta T$ directly from the magnetocaloric measurements. The time to cross the transition at $H_{\mathrm{m}}$ is $\sim 0.6\,\mathrm{ms}$ - significantly shorter than the thermal relaxation timescale $\tau$ of the sample in the FP state which is around $10\,\mathrm{ms}$ for our equipment. $\tau$ was estimated from the $T(t)$ behavior above $H_{\mathrm{m}}$ [Fig. \ref{fig:figure5}(a)], yet it is clearly longer in the SC$_{\rm FP}$ region of the phase diagram. We obtain the reversible temperature changes at the metamagnetic transition through $\Delta T_{\mathrm{rev}} = (\Delta T_{\mathrm{FPup}} - \Delta T_{\mathrm{FPdown}})/2$, where the subscripts ``up'' and ``down'' refer to the up- and downsweeps of the field respectively. On the other hand, irreversible processes such as Joule heating contribute to the temperature change in both field-sweep directions, therefore $\Delta T_{\mathrm{irr}} = (\Delta T_{\mathrm{FPup}} + \Delta T_{\mathrm{FPdown}})/2$.

\begin{figure}
\includegraphics{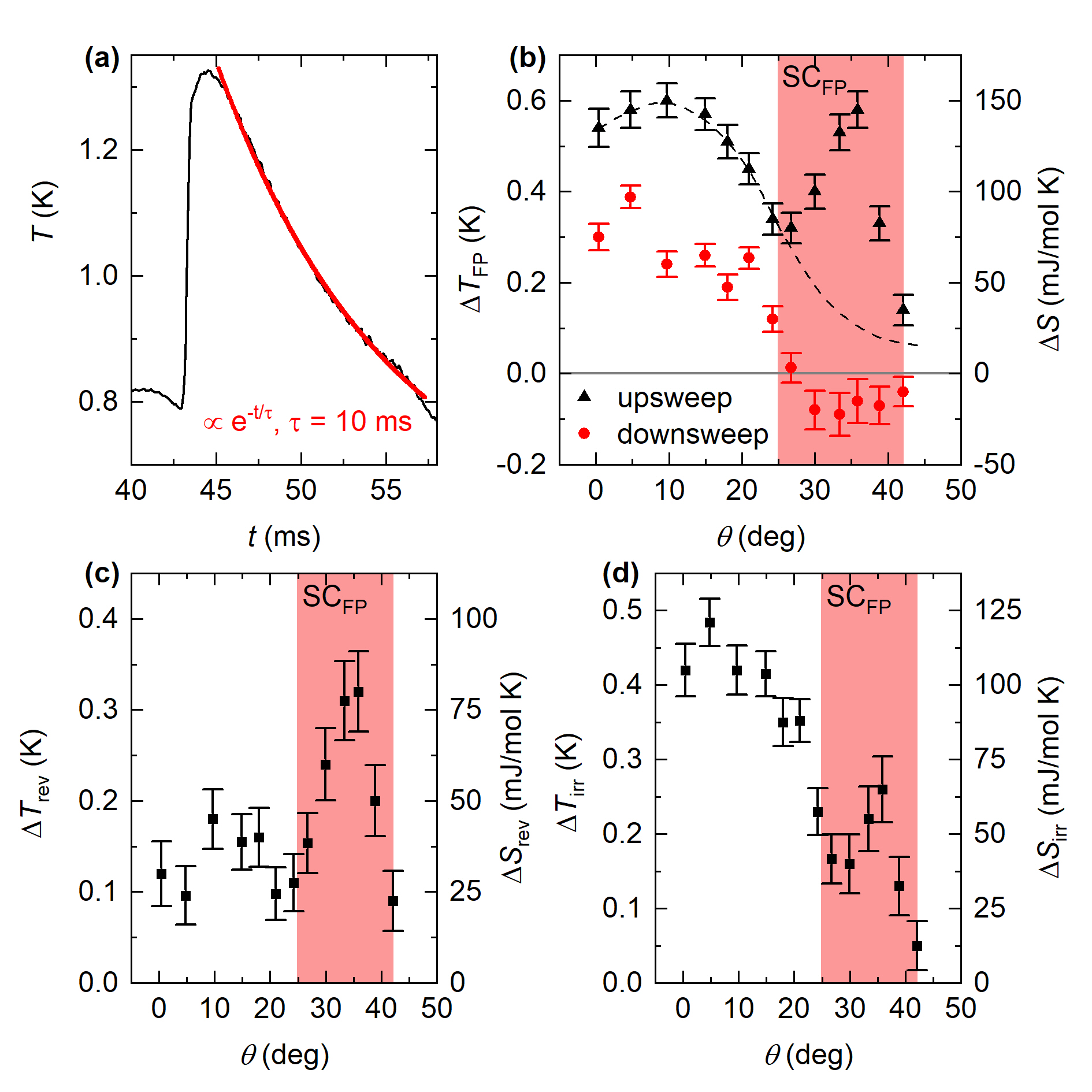}
\caption{(a) Temperature vs. time during the up-sweep of the magnetic field pulse for $H\parallel b$. The time frame shows the metamagnetic transition and the subsequent relaxation back to the bath temperature, which is approximated by an exponential decay (red line). (b) Temperature change $\Delta T_{\mathrm{FP}}(\theta)$ at the metamagnetic transition during the up-sweep (black triangles) and down-sweep (red circles) of the magnetic field. (c) Reversible and (d) irreversible component of $\Delta T_{\mathrm{FP}}$ as a function of the angle $\theta$ (left axes). While reversible processes are prevalent in the SC$_{\rm FP}$ phase at $\theta \simeq 33^o$, irreversible mechanisms or dissipation dominate in the small $\theta$ region.The corresponding entropy changes are shown on the right axes of each figure.}
\label{fig:figure5}
\end{figure}

Using that $C_{p}/T \approx 250$~mJmol$^{-1}$K$^{-2}$ and assuming that $C_{p}$ shows weak temperature dependence below $2\,\mathrm{K}$ at $35\,\mathrm{T}$ \cite{imajo_thermodynamic_2019, rosuel_2023}, for $\theta <25^\circ$ we obtain an almost constant value, $\Delta S_{\mathrm{rev}} \approx 30$~mJmol$^{-1}$K$^{-1}$. Within the SC$_{\rm FP}$ phase, $\Delta S_{\rm rev}$ increases, peaking at $\Delta S_{\mathrm{rev}} \approx 80$~mJmol$^{-1}$K$^{-1}$ close to $\theta=35^\circ$ [Fig.~\ref{fig:figure5}(c)]~\footnote{We note a discrepancy in the entropy change at the metamagnetic transition between the present results and the values reported in the brief report by Imajo \textit{et al.} Ref.~\cite{imajo_thermodynamic_2019}. The difference is likely related to the conditions in the present work, \textit{i.e.} a stronger link to the thermal bath needed to reach lower $^3$He temperatures leading to the quasi-adiabatic magnetization of the sample. The difference does not affect the conclusions of the current paper.} Therefore, entering the SC$_{\rm FP}$ phase releases an additional $\approx 50$~mJmol$^{-1}$K$^{-1}$ in latent heat. Assuming (as justified above) that the SC$_{\rm FP}$ represents a field-induced superconducting state, the additional latent heat is likely to result from the formation of a gap at the Fermi energy and an entropy reduction due to pair condensation~\cite{aoki_unconventional_2022}. 

The irreversible component $\Delta S_{irr}$ mainly consists of hysteretic losses during the first-order metamagnetic transition and, bearing in mind the similarity of the behavior of the PDO data in the SC$_{\rm FP}$ state to that in the SC$_{\rm PM}$ phase (see also Ref.~\cite{ran_extreme_2019}), what is likely to be dissipation due to vortex movement. As shown in Fig.~\ref{fig:figure5}(d), rotating $H$ to higher $\theta$ leads to an overall decrease in $\Delta S_{\mathrm{irr}}$, apart from a small local uptick around $\theta = 35^{\circ}$. As this is roughly in the middle of the $\theta$ range over which the SC$_{\rm FP}$ phase occurs, it possibly coincides with dissipation caused by a combination of metamagnetism, some vortex motion, and lack of perfect adiabaticity. Note that while $H_{\mathrm{m}}$ increases with increasing $\theta$, the jump in the magnetization at $H_{\mathrm{m}}$ at $1.4\,\mathrm{K}$ does not change significantly between $H\parallel b$ and $H\parallel [011]$ \cite{miyake_enhancement_2021}. Torque measurements shown in Fig. \ref{fig:figure6} also vary smoothly as a function of angle. Therefore it is unlikely that the small irreversible heat involved when entering the SC$_{\rm FP}$ phase is of magnetic origin.

Finally, we remark that the boundaries between the various low-temperature/high-magnetic-field phases of UTe$_2$ derived in this work from PDO and MCE data match those for CVT-grown samples in the literature~\cite{ran_extreme_2019, aoki_unconventional_2022} very closely. This is of interest because the zero- or low-field behaviour of UTe$_2$ seems very sensitive to the source, growth method, and quality of the crystals used (an excellent summary is given in Ref.~\cite{aoki_unconventional_2022}). The present study employs crystals from different sources to those used to produce the phase diagrams reported in Refs.~\cite{ran_extreme_2019, aoki_unconventional_2022}, perhaps suggesting that the high-field properties of UTe$_2$ are less sensitive to sample dependent disorder than those in zero or small magnetic fields \cite{wu_arxiv_2023, frank_arxiv_2023}.

\section*{Conclusions}

The simultaneous zero-electric current magnetocaloric effect, MHz conductivity measurements, and angular dependent torque magnetometry are carried out on single crystals of UTe$_2$ as a function of magnetic field magnitude and orientation, using pulsed magnetic fields of up to 55 T. A pronounced and fully reversible magnetocaloric effect characteristic of a thermally decoupled (adiabatic) state is observed close to the metamagnetic transition into the proposed high field SC$_{\rm FP}$ phase. This amounts to compelling evidence for the stabilization of a field-induced energy-gaped state of concurrent high electrical- and poor thermal conductivity, {\it i.e.,} the first thermodynamic evidence that the SC$_{\rm FP}$ state represents a field-stabilized bulk superconducting phase of UTe$_2$, of likely node-less order parameter. Additionally, with the magnetic field aligned close to the $b$-axis, a more subtle feature is observed around $15\,\mathrm{T}$, supporting the notion that the superconducting SC$_{\rm PM}$ and SC$_{\rm RE}$ states represent separate, distinct phases.

\section*{Experimental details}

Single crystals of UTe$_2$ are grown using chemical vapor transport; the conditions are the same as for sample s4 described in Ref. \cite{rosa_single-component_2021}, where further details can be found. To provide initial characterization prior to the pulsed-field experiments, heat-capacity measurements are performed using a commercial calorimeter that utilizes a quasi-adiabatic thermal relaxation technique. In addition, the electrical resistivity $\rho$ is characterized using a standard four-probe configuration with an AC resistance bridge. Resistivity (not shown) and heat-capacity measurements on crystals from this batch show a single sharp transition around $1.9\,\mathrm{K}$ [Fig. \ref{fig:figure1}(b)].

Fig.~\ref{fig:figure1}(c) shows a schematic drawing of the sample environment for the pulsed-field experiments. The pancake coil for the PDO measurements (10 turns of insulated 50-gauge copper wire) is sandwiched between a G10 holder and the single-crystal UTe$_2$ sample. The sample was coated with a thin film of GE varnish to avoid electrical contact with the layers above. The MCE thermometer is an approximately $100\,\mathrm{nm}$ thick semiconducting AuGe film ($16\,\mathrm{at}\%$ Au) deposited directly on the varnish-coated sample to ensure good thermal coupling between sample and film. To improve the contact resistance, Au pads are deposited on the AuGe film. The AuGe film is calibrated against a commercial Cernox sensor; film resistances range from $6\,\Omega$ at room temperature to $250\,\Omega$ at $0.6\,\mathrm{K}$. The sample is glued to the holder with Stycast\textregistered ~epoxy to prevent any sample movement due to the large magnetic torque when the field is aligned close to the $b$-axis. 

The PDO measurements employ equipment similar to that described in Refs.~\cite{altarawneh_proximity_2009, ghannadzadeh_measurement_2011,Xiang2021,PCHo,SMYLIE,NIKOLO}; the technique is well established for mapping the irreversibility and upper critical fields of superconductors in pulsed magnetic fields~\cite{SMYLIE,NIKOLO}. The magnetocaloric and PDO experiments were performed in the NHMFL's mid-pulse magnet, which provides a peak magnetic field of $55\,\mathrm{T}$ with a rise time of approximately $30\,\mathrm{ms}$ and a total pulse duration of $500\,\mathrm{ms}$. A typical field pulse and its derivative are shown in Fig. \ref{fig:figure1}(d). The sample holder was fixed to the rotating platform of a cryogenic goniometer~\cite{goniometer} placed within a $^3$He cryostat. The sample was immersed in liquid ${}^{3}\mathrm{He}$ at a bath temperature of $0.6\pm0.1$~K during the field pulses. 

Additionally, we conducted piezo torque magnetometry measurements in pulsed magnetic fields up to $75\,\mathrm{T}$ by using membrane-type surface-stress sensors at the NHMFL at LANL with a high-frequency ($\approx 300\,\mathrm{kHz}$) AC excitation current of $\approx 500 \,\mu\mathrm{A}$. The angular dependent torque measurements were performed at $0.7\,\mathrm{K}$ with the sample immersed in liquid ${}^{3}$He. In the experiments, we used a balanced Wheatstone bridge between the piezoresistive pathways. Crystals were mounted with the $b$ axis perpendicular to the cantilever plane.}

\section*{Acknowledgements}
We thank M. Lee, L. Civale, and A. Shehter for insightful discussions. A portion of this work was performed at the National High Magnetic Field Laboratory, which is supported by the NSF Cooperative Agreement No. DMR-1644779, the U.S. DOE and the State of Florida. This material is based upon work supported by the U.S. Department of Energy, Office of Science, National Quantum Information Science Research Centers. S.M.T acknowledges support from the Los Alamos Laboratory Directed Research and Development program through project 20210064DR. J.S. and M.J. thank the DoE BES FWP ``Science of 100~T'' for support in developing techniques used in these experiments. R.S., Y. L., and M.J. acknowledge support by the NHMFL UCGP program and the G. T. Seaborg Institute Postdoctoral Fellow Program under project number 20210527CR.

\section*{Apendix A: Complete angular dependent Magnetocaloric data set}

\begin{figure*}\includegraphics{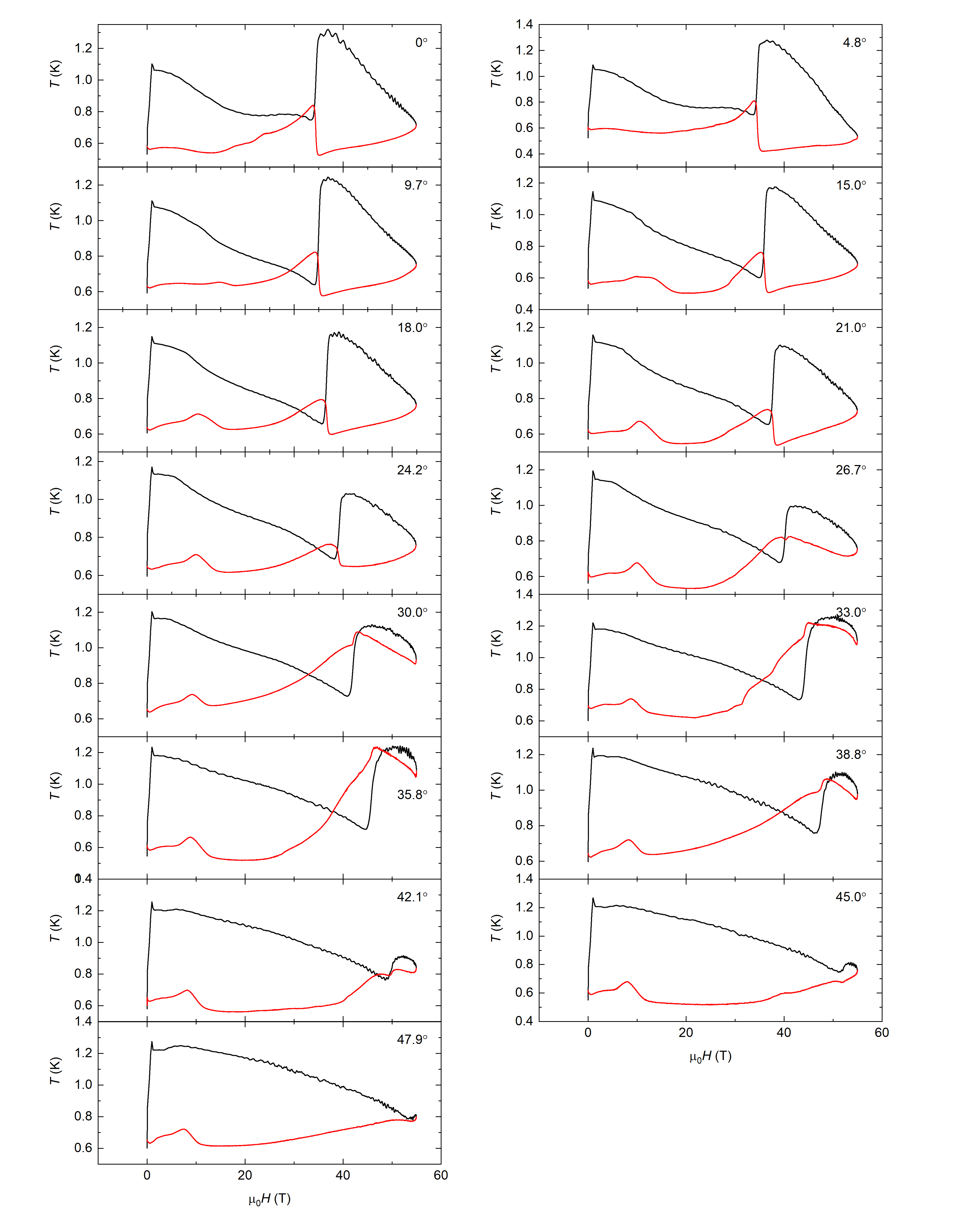}[h]
\caption{Sample temperature vs. magnetic field for different angles denoted in the graphs, where $0^{\circ}$ is $H\parallel b$ and $90^{\circ}$ is $H\parallel c$. Field up- and down-sweep data are depicted as black and red lines respectively.}
\label{fig:figure7}
\end{figure*}

Here we show the entire angular dependent magnetocaloric data set (Fig. \ref{fig:figure7}) measured with the sample immersed in liquid ${}^{3}$He. The data was used to generate the contour plots shown in Fig \ref{fig:figure4}.

\newpage
\bibliography{UTe2_MCE}

\begin{thebibliography}{43}%
\makeatletter
\providecommand \@ifxundefined [1]{%
 \@ifx{#1\undefined}
}%
\providecommand \@ifnum [1]{%
 \ifnum #1\expandafter \@firstoftwo
 \else \expandafter \@secondoftwo
 \fi
}%
\providecommand \@ifx [1]{%
 \ifx #1\expandafter \@firstoftwo
 \else \expandafter \@secondoftwo
 \fi
}%
\providecommand \natexlab [1]{#1}%
\providecommand \enquote  [1]{``#1''}%
\providecommand \bibnamefont  [1]{#1}%
\providecommand \bibfnamefont [1]{#1}%
\providecommand \citenamefont [1]{#1}%
\providecommand \href@noop [0]{\@secondoftwo}%
\providecommand \href [0]{\begingroup \@sanitize@url \@href}%
\providecommand \@href[1]{\@@startlink{#1}\@@href}%
\providecommand \@@href[1]{\endgroup#1\@@endlink}%
\providecommand \@sanitize@url [0]{\catcode `\\12\catcode `\$12\catcode
  `\&12\catcode `\#12\catcode `\^12\catcode `\_12\catcode `\%12\relax}%
\providecommand \@@startlink[1]{}%
\providecommand \@@endlink[0]{}%
\providecommand \url  [0]{\begingroup\@sanitize@url \@url }%
\providecommand \@url [1]{\endgroup\@href {#1}{\urlprefix }}%
\providecommand \urlprefix  [0]{URL }%
\providecommand \Eprint [0]{\href }%
\providecommand \doibase [0]{http://dx.doi.org/}%
\providecommand \selectlanguage [0]{\@gobble}%
\providecommand \bibinfo  [0]{\@secondoftwo}%
\providecommand \bibfield  [0]{\@secondoftwo}%
\providecommand \translation [1]{[#1]}%
\providecommand \BibitemOpen [0]{}%
\providecommand \bibitemStop [0]{}%
\providecommand \bibitemNoStop [0]{.\EOS\space}%
\providecommand \EOS [0]{\spacefactor3000\relax}%
\providecommand \BibitemShut  [1]{\csname bibitem#1\endcsname}%
\let\auto@bib@innerbib\@empty
\bibitem [{\citenamefont {Ran}\ \emph {et~al.}(2019{\natexlab{a}})\citenamefont
  {Ran}, \citenamefont {Eckberg}, \citenamefont {Ding}, \citenamefont
  {Furukawa}, \citenamefont {Metz}, \citenamefont {Saha}, \citenamefont {Liu},
  \citenamefont {Zic}, \citenamefont {Kim}, \citenamefont {Paglione},\ and\
  \citenamefont {Butch}}]{ran_nearly_2019}%
  \BibitemOpen
  \bibfield  {author} {\bibinfo {author} {\bibfnamefont {S.}~\bibnamefont
  {Ran}}, \bibinfo {author} {\bibfnamefont {C.}~\bibnamefont {Eckberg}},
  \bibinfo {author} {\bibfnamefont {Q.-P.}\ \bibnamefont {Ding}}, \bibinfo
  {author} {\bibfnamefont {Y.}~\bibnamefont {Furukawa}}, \bibinfo {author}
  {\bibfnamefont {T.}~\bibnamefont {Metz}}, \bibinfo {author} {\bibfnamefont
  {S.~R.}\ \bibnamefont {Saha}}, \bibinfo {author} {\bibfnamefont {I.-L.}\
  \bibnamefont {Liu}}, \bibinfo {author} {\bibfnamefont {M.}~\bibnamefont
  {Zic}}, \bibinfo {author} {\bibfnamefont {H.}~\bibnamefont {Kim}}, \bibinfo
  {author} {\bibfnamefont {J.}~\bibnamefont {Paglione}}, \ and\ \bibinfo
  {author} {\bibfnamefont {N.~P.}\ \bibnamefont {Butch}},\ }\href {\doibase
  10.1126/science.aav8645} {\bibfield  {journal} {\bibinfo  {journal}
  {Science}\ }\textbf {\bibinfo {volume} {365}},\ \bibinfo {pages} {684}
  (\bibinfo {year} {2019}{\natexlab{a}})}\BibitemShut {NoStop}%
\bibitem [{\citenamefont {Ran}\ \emph {et~al.}(2019{\natexlab{b}})\citenamefont
  {Ran}, \citenamefont {Liu}, \citenamefont {Eo}, \citenamefont {Campbell},
  \citenamefont {Neves}, \citenamefont {Fuhrman}, \citenamefont {Saha},
  \citenamefont {Eckberg}, \citenamefont {Kim}, \citenamefont {Graf},
  \citenamefont {Balakirev}, \citenamefont {Singleton}, \citenamefont
  {Paglione},\ and\ \citenamefont {Butch}}]{ran_extreme_2019}%
  \BibitemOpen
  \bibfield  {author} {\bibinfo {author} {\bibfnamefont {S.}~\bibnamefont
  {Ran}}, \bibinfo {author} {\bibfnamefont {I.-L.}\ \bibnamefont {Liu}},
  \bibinfo {author} {\bibfnamefont {Y.~S.}\ \bibnamefont {Eo}}, \bibinfo
  {author} {\bibfnamefont {D.~J.}\ \bibnamefont {Campbell}}, \bibinfo {author}
  {\bibfnamefont {P.~M.}\ \bibnamefont {Neves}}, \bibinfo {author}
  {\bibfnamefont {W.~T.}\ \bibnamefont {Fuhrman}}, \bibinfo {author}
  {\bibfnamefont {S.~R.}\ \bibnamefont {Saha}}, \bibinfo {author}
  {\bibfnamefont {C.}~\bibnamefont {Eckberg}}, \bibinfo {author} {\bibfnamefont
  {H.}~\bibnamefont {Kim}}, \bibinfo {author} {\bibfnamefont {D.}~\bibnamefont
  {Graf}}, \bibinfo {author} {\bibfnamefont {F.}~\bibnamefont {Balakirev}},
  \bibinfo {author} {\bibfnamefont {J.}~\bibnamefont {Singleton}}, \bibinfo
  {author} {\bibfnamefont {J.}~\bibnamefont {Paglione}}, \ and\ \bibinfo
  {author} {\bibfnamefont {N.~P.}\ \bibnamefont {Butch}},\ }\href {\doibase
  10.1038/s41567-019-0670-x} {\bibfield  {journal} {\bibinfo  {journal} {Nature
  Physics}\ }\textbf {\bibinfo {volume} {15}},\ \bibinfo {pages} {1250}
  (\bibinfo {year} {2019}{\natexlab{b}})}\BibitemShut {NoStop}%
\bibitem [{\citenamefont {Knafo}\ \emph {et~al.}(2019)\citenamefont {Knafo},
  \citenamefont {Vališka}, \citenamefont {Braithwaite}, \citenamefont
  {Lapertot}, \citenamefont {Knebel}, \citenamefont {Pourret}, \citenamefont
  {Brison}, \citenamefont {Flouquet},\ and\ \citenamefont
  {Aoki}}]{knafo_magnetic-field-induced_2019}%
  \BibitemOpen
  \bibfield  {author} {\bibinfo {author} {\bibfnamefont {W.}~\bibnamefont
  {Knafo}}, \bibinfo {author} {\bibfnamefont {M.}~\bibnamefont {Vališka}},
  \bibinfo {author} {\bibfnamefont {D.}~\bibnamefont {Braithwaite}}, \bibinfo
  {author} {\bibfnamefont {G.}~\bibnamefont {Lapertot}}, \bibinfo {author}
  {\bibfnamefont {G.}~\bibnamefont {Knebel}}, \bibinfo {author} {\bibfnamefont
  {A.}~\bibnamefont {Pourret}}, \bibinfo {author} {\bibfnamefont {J.-P.}\
  \bibnamefont {Brison}}, \bibinfo {author} {\bibfnamefont {J.}~\bibnamefont
  {Flouquet}}, \ and\ \bibinfo {author} {\bibfnamefont {D.}~\bibnamefont
  {Aoki}},\ }\href {\doibase 10.7566/JPSJ.88.063705} {\bibfield  {journal}
  {\bibinfo  {journal} {Journal of the Physical Society of Japan}\ }\textbf
  {\bibinfo {volume} {88}},\ \bibinfo {pages} {063705} (\bibinfo {year}
  {2019})}\BibitemShut {NoStop}%
\bibitem [{\citenamefont {Knebel}\ \emph {et~al.}(2019)\citenamefont {Knebel},
  \citenamefont {Knafo}, \citenamefont {Pourret}, \citenamefont {Niu},
  \citenamefont {Vališka}, \citenamefont {Braithwaite}, \citenamefont
  {Lapertot}, \citenamefont {Nardone}, \citenamefont {Zitouni}, \citenamefont
  {Mishra}, \citenamefont {Sheikin}, \citenamefont {Seyfarth}, \citenamefont
  {Brison}, \citenamefont {Aoki},\ and\ \citenamefont
  {Flouquet}}]{knebel_field-reentrant_2019}%
  \BibitemOpen
  \bibfield  {author} {\bibinfo {author} {\bibfnamefont {G.}~\bibnamefont
  {Knebel}}, \bibinfo {author} {\bibfnamefont {W.}~\bibnamefont {Knafo}},
  \bibinfo {author} {\bibfnamefont {A.}~\bibnamefont {Pourret}}, \bibinfo
  {author} {\bibfnamefont {Q.}~\bibnamefont {Niu}}, \bibinfo {author}
  {\bibfnamefont {M.}~\bibnamefont {Vališka}}, \bibinfo {author}
  {\bibfnamefont {D.}~\bibnamefont {Braithwaite}}, \bibinfo {author}
  {\bibfnamefont {G.}~\bibnamefont {Lapertot}}, \bibinfo {author}
  {\bibfnamefont {M.}~\bibnamefont {Nardone}}, \bibinfo {author} {\bibfnamefont
  {A.}~\bibnamefont {Zitouni}}, \bibinfo {author} {\bibfnamefont
  {S.}~\bibnamefont {Mishra}}, \bibinfo {author} {\bibfnamefont
  {I.}~\bibnamefont {Sheikin}}, \bibinfo {author} {\bibfnamefont
  {G.}~\bibnamefont {Seyfarth}}, \bibinfo {author} {\bibfnamefont {J.-P.}\
  \bibnamefont {Brison}}, \bibinfo {author} {\bibfnamefont {D.}~\bibnamefont
  {Aoki}}, \ and\ \bibinfo {author} {\bibfnamefont {J.}~\bibnamefont
  {Flouquet}},\ }\href {\doibase 10.7566/JPSJ.88.063707} {\bibfield  {journal}
  {\bibinfo  {journal} {Journal of the Physical Society of Japan}\ }\textbf
  {\bibinfo {volume} {88}},\ \bibinfo {pages} {063707} (\bibinfo {year}
  {2019})}\BibitemShut {NoStop}%
\bibitem [{\citenamefont {Nakamine}\ \emph {et~al.}(2019)\citenamefont
  {Nakamine}, \citenamefont {Kitagawa}, \citenamefont {Ishida}, \citenamefont
  {Tokunaga}, \citenamefont {Sakai}, \citenamefont {Kambe}, \citenamefont
  {Nakamura}, \citenamefont {Shimizu}, \citenamefont {Homma}, \citenamefont
  {Li}, \citenamefont {Honda},\ and\ \citenamefont
  {Aoki}}]{nakamine_superconducting_2019}%
  \BibitemOpen
  \bibfield  {author} {\bibinfo {author} {\bibfnamefont {G.}~\bibnamefont
  {Nakamine}}, \bibinfo {author} {\bibfnamefont {S.}~\bibnamefont {Kitagawa}},
  \bibinfo {author} {\bibfnamefont {K.}~\bibnamefont {Ishida}}, \bibinfo
  {author} {\bibfnamefont {Y.}~\bibnamefont {Tokunaga}}, \bibinfo {author}
  {\bibfnamefont {H.}~\bibnamefont {Sakai}}, \bibinfo {author} {\bibfnamefont
  {S.}~\bibnamefont {Kambe}}, \bibinfo {author} {\bibfnamefont
  {A.}~\bibnamefont {Nakamura}}, \bibinfo {author} {\bibfnamefont
  {Y.}~\bibnamefont {Shimizu}}, \bibinfo {author} {\bibfnamefont
  {Y.}~\bibnamefont {Homma}}, \bibinfo {author} {\bibfnamefont
  {D.}~\bibnamefont {Li}}, \bibinfo {author} {\bibfnamefont {F.}~\bibnamefont
  {Honda}}, \ and\ \bibinfo {author} {\bibfnamefont {D.}~\bibnamefont {Aoki}},\
  }\href {\doibase 10.7566/JPSJ.88.113703} {\bibfield  {journal} {\bibinfo
  {journal} {Journal of the Physical Society of Japan}\ }\textbf {\bibinfo
  {volume} {88}},\ \bibinfo {pages} {113703} (\bibinfo {year}
  {2019})}\BibitemShut {NoStop}%
\bibitem [{\citenamefont {Jiao}\ \emph {et~al.}(2020)\citenamefont {Jiao},
  \citenamefont {Howard}, \citenamefont {Ran}, \citenamefont {Wang},
  \citenamefont {Rodriguez}, \citenamefont {Sigrist}, \citenamefont {Wang},
  \citenamefont {Butch},\ and\ \citenamefont {Madhavan}}]{jiao_stm_2020}%
  \BibitemOpen
  \bibfield  {author} {\bibinfo {author} {\bibfnamefont {L.}~\bibnamefont
  {Jiao}}, \bibinfo {author} {\bibfnamefont {S.}~\bibnamefont {Howard}},
  \bibinfo {author} {\bibfnamefont {S.}~\bibnamefont {Ran}}, \bibinfo {author}
  {\bibfnamefont {Z.}~\bibnamefont {Wang}}, \bibinfo {author} {\bibfnamefont
  {J.~O.}\ \bibnamefont {Rodriguez}}, \bibinfo {author} {\bibfnamefont
  {M.}~\bibnamefont {Sigrist}}, \bibinfo {author} {\bibfnamefont
  {Z.}~\bibnamefont {Wang}}, \bibinfo {author} {\bibfnamefont {N.~P.}\
  \bibnamefont {Butch}}, \ and\ \bibinfo {author} {\bibfnamefont
  {V.}~\bibnamefont {Madhavan}},\ }\href {\doibase 10.1038/s41586-020-2122-2}
  {\bibfield  {journal} {\bibinfo  {journal} {Nature}\ }\textbf {\bibinfo
  {volume} {579}},\ \bibinfo {pages} {523} (\bibinfo {year}
  {2020})}\BibitemShut {NoStop}%
\bibitem [{\citenamefont {Aoki}\ \emph {et~al.}(2022)\citenamefont {Aoki},
  \citenamefont {Brison}, \citenamefont {Flouquet}, \citenamefont {Ishida},
  \citenamefont {Knebel}, \citenamefont {Tokunaga},\ and\ \citenamefont
  {Yanase}}]{aoki_unconventional_2022}%
  \BibitemOpen
  \bibfield  {author} {\bibinfo {author} {\bibfnamefont {D.}~\bibnamefont
  {Aoki}}, \bibinfo {author} {\bibfnamefont {J.-P.}\ \bibnamefont {Brison}},
  \bibinfo {author} {\bibfnamefont {J.}~\bibnamefont {Flouquet}}, \bibinfo
  {author} {\bibfnamefont {K.}~\bibnamefont {Ishida}}, \bibinfo {author}
  {\bibfnamefont {G.}~\bibnamefont {Knebel}}, \bibinfo {author} {\bibfnamefont
  {Y.}~\bibnamefont {Tokunaga}}, \ and\ \bibinfo {author} {\bibfnamefont
  {Y.}~\bibnamefont {Yanase}},\ }\href {\doibase 10.1088/1361-648x/ac5863}
  {\bibfield  {journal} {\bibinfo  {journal} {Journal of Physics: Condensed
  Matter}\ }\textbf {\bibinfo {volume} {34}},\ \bibinfo {pages} {243002}
  (\bibinfo {year} {2022})}\BibitemShut {NoStop}%
\bibitem [{\citenamefont {Rosuel}\ \emph {et~al.}(2023)\citenamefont {Rosuel},
  \citenamefont {Marcenat}, \citenamefont {Knebel}, \citenamefont {Klein},
  \citenamefont {Pourret}, \citenamefont {Marquardt}, \citenamefont {Niu},
  \citenamefont {Rousseau}, \citenamefont {Demuer}, \citenamefont {Seyfarth},
  \citenamefont {Lapertot}, \citenamefont {Aoki}, \citenamefont {Braithwaite},
  \citenamefont {Flouquet},\ and\ \citenamefont {Brison}}]{rosuel_2023}%
  \BibitemOpen
  \bibfield  {author} {\bibinfo {author} {\bibfnamefont {A.}~\bibnamefont
  {Rosuel}}, \bibinfo {author} {\bibfnamefont {C.}~\bibnamefont {Marcenat}},
  \bibinfo {author} {\bibfnamefont {G.}~\bibnamefont {Knebel}}, \bibinfo
  {author} {\bibfnamefont {T.}~\bibnamefont {Klein}}, \bibinfo {author}
  {\bibfnamefont {A.}~\bibnamefont {Pourret}}, \bibinfo {author} {\bibfnamefont
  {N.}~\bibnamefont {Marquardt}}, \bibinfo {author} {\bibfnamefont
  {Q.}~\bibnamefont {Niu}}, \bibinfo {author} {\bibfnamefont {S.}~\bibnamefont
  {Rousseau}}, \bibinfo {author} {\bibfnamefont {A.}~\bibnamefont {Demuer}},
  \bibinfo {author} {\bibfnamefont {G.}~\bibnamefont {Seyfarth}}, \bibinfo
  {author} {\bibfnamefont {G.}~\bibnamefont {Lapertot}}, \bibinfo {author}
  {\bibfnamefont {D.}~\bibnamefont {Aoki}}, \bibinfo {author} {\bibfnamefont
  {D.}~\bibnamefont {Braithwaite}}, \bibinfo {author} {\bibfnamefont
  {J.}~\bibnamefont {Flouquet}}, \ and\ \bibinfo {author} {\bibfnamefont
  {J.~P.}\ \bibnamefont {Brison}},\ }\href {\doibase
  10.1103/PhysRevX.13.011022} {\bibfield  {journal} {\bibinfo  {journal} {Phys.
  Rev. X}\ }\textbf {\bibinfo {volume} {13}},\ \bibinfo {pages} {011022}
  (\bibinfo {year} {2023})}\BibitemShut {NoStop}%
\bibitem [{\citenamefont {Matsumura}\ \emph {et~al.}(2023)\citenamefont
  {Matsumura}, \citenamefont {Fujibayashi}, \citenamefont {Kinjo},
  \citenamefont {Kitagawa}, \citenamefont {Ishida}, \citenamefont {Tokunaga},
  \citenamefont {Sakai}, \citenamefont {Kambe}, \citenamefont {Nakamura},
  \citenamefont {Shimizu}, \citenamefont {Homma}, \citenamefont {Honda}, ,\
  and\ \citenamefont {Aoki}}]{matsumura_2023}%
  \BibitemOpen
  \bibfield  {author} {\bibinfo {author} {\bibfnamefont {H.}~\bibnamefont
  {Matsumura}}, \bibinfo {author} {\bibfnamefont {H.}~\bibnamefont
  {Fujibayashi}}, \bibinfo {author} {\bibfnamefont {K.}~\bibnamefont {Kinjo}},
  \bibinfo {author} {\bibfnamefont {S.}~\bibnamefont {Kitagawa}}, \bibinfo
  {author} {\bibfnamefont {K.}~\bibnamefont {Ishida}}, \bibinfo {author}
  {\bibfnamefont {Y.}~\bibnamefont {Tokunaga}}, \bibinfo {author}
  {\bibfnamefont {H.}~\bibnamefont {Sakai}}, \bibinfo {author} {\bibfnamefont
  {S.}~\bibnamefont {Kambe}}, \bibinfo {author} {\bibfnamefont
  {A.}~\bibnamefont {Nakamura}}, \bibinfo {author} {\bibfnamefont
  {Y.}~\bibnamefont {Shimizu}}, \bibinfo {author} {\bibfnamefont
  {D.}~\bibnamefont {Homma}, \bibfnamefont {Y.~Li}}, \bibinfo {author}
  {\bibfnamefont {F.}~\bibnamefont {Honda}}, , \ and\ \bibinfo {author}
  {\bibfnamefont {D.}~\bibnamefont {Aoki}},\ }\href {\doibase
  https://doi.org/10.7566/JPSJ.92.063701} {\bibfield  {journal} {\bibinfo
  {journal} {Jou. Phys. Soc. Jpn.}\ }\textbf {\bibinfo {volume} {92}},\
  \bibinfo {pages} {063701} (\bibinfo {year} {2023})}\BibitemShut {NoStop}%
\bibitem [{\citenamefont {Sundar}\ \emph {et~al.}(2019)\citenamefont {Sundar},
  \citenamefont {Gheidi}, \citenamefont {Akintola}, \citenamefont {Cote},
  \citenamefont {Dunsiger}, \citenamefont {Ran}, \citenamefont {Butch},
  \citenamefont {Saha}, \citenamefont {Paglione},\ and\ \citenamefont
  {Sonier}}]{sundar_coexistence_2019}%
  \BibitemOpen
  \bibfield  {author} {\bibinfo {author} {\bibfnamefont {S.}~\bibnamefont
  {Sundar}}, \bibinfo {author} {\bibfnamefont {S.}~\bibnamefont {Gheidi}},
  \bibinfo {author} {\bibfnamefont {K.}~\bibnamefont {Akintola}}, \bibinfo
  {author} {\bibfnamefont {A.~M.}\ \bibnamefont {Cote}}, \bibinfo {author}
  {\bibfnamefont {S.~R.}\ \bibnamefont {Dunsiger}}, \bibinfo {author}
  {\bibfnamefont {S.}~\bibnamefont {Ran}}, \bibinfo {author} {\bibfnamefont
  {N.~P.}\ \bibnamefont {Butch}}, \bibinfo {author} {\bibfnamefont {S.~R.}\
  \bibnamefont {Saha}}, \bibinfo {author} {\bibfnamefont {J.}~\bibnamefont
  {Paglione}}, \ and\ \bibinfo {author} {\bibfnamefont {J.~E.}\ \bibnamefont
  {Sonier}},\ }\href {\doibase 10.1103/PhysRevB.100.140502} {\bibfield
  {journal} {\bibinfo  {journal} {Physical Review B}\ }\textbf {\bibinfo
  {volume} {100}},\ \bibinfo {pages} {140502(R)} (\bibinfo {year}
  {2019})}\BibitemShut {NoStop}%
\bibitem [{\citenamefont {Paulsen}\ \emph {et~al.}(2021)\citenamefont
  {Paulsen}, \citenamefont {Knebel}, \citenamefont {Lapertot}, \citenamefont
  {Braithwaite}, \citenamefont {Pourret}, \citenamefont {Aoki}, \citenamefont
  {Hardy}, \citenamefont {Flouquet},\ and\ \citenamefont
  {Brison}}]{paulsen_anomalous_2021}%
  \BibitemOpen
  \bibfield  {author} {\bibinfo {author} {\bibfnamefont {C.}~\bibnamefont
  {Paulsen}}, \bibinfo {author} {\bibfnamefont {G.}~\bibnamefont {Knebel}},
  \bibinfo {author} {\bibfnamefont {G.}~\bibnamefont {Lapertot}}, \bibinfo
  {author} {\bibfnamefont {D.}~\bibnamefont {Braithwaite}}, \bibinfo {author}
  {\bibfnamefont {A.}~\bibnamefont {Pourret}}, \bibinfo {author} {\bibfnamefont
  {D.}~\bibnamefont {Aoki}}, \bibinfo {author} {\bibfnamefont {F.}~\bibnamefont
  {Hardy}}, \bibinfo {author} {\bibfnamefont {J.}~\bibnamefont {Flouquet}}, \
  and\ \bibinfo {author} {\bibfnamefont {J.-P.}\ \bibnamefont {Brison}},\
  }\href {\doibase 10.1103/PhysRevB.103.L180501} {\bibfield  {journal}
  {\bibinfo  {journal} {Physical Review B}\ }\textbf {\bibinfo {volume}
  {103}},\ \bibinfo {pages} {L180501} (\bibinfo {year} {2021})}\BibitemShut
  {NoStop}%
\bibitem [{\citenamefont {Duan}\ \emph {et~al.}(2020)\citenamefont {Duan},
  \citenamefont {Sasmal}, \citenamefont {Maple}, \citenamefont {Podlesnyak},
  \citenamefont {Zhu}, \citenamefont {Si},\ and\ \citenamefont
  {Dai}}]{duan_incommensurate_2020}%
  \BibitemOpen
  \bibfield  {author} {\bibinfo {author} {\bibfnamefont {C.}~\bibnamefont
  {Duan}}, \bibinfo {author} {\bibfnamefont {K.}~\bibnamefont {Sasmal}},
  \bibinfo {author} {\bibfnamefont {M.~B.}\ \bibnamefont {Maple}}, \bibinfo
  {author} {\bibfnamefont {A.}~\bibnamefont {Podlesnyak}}, \bibinfo {author}
  {\bibfnamefont {J.-X.}\ \bibnamefont {Zhu}}, \bibinfo {author} {\bibfnamefont
  {Q.}~\bibnamefont {Si}}, \ and\ \bibinfo {author} {\bibfnamefont
  {P.}~\bibnamefont {Dai}},\ }\href {\doibase 10.1103/PhysRevLett.125.237003}
  {\bibfield  {journal} {\bibinfo  {journal} {Physical Review Letters}\
  }\textbf {\bibinfo {volume} {125}},\ \bibinfo {pages} {237003} (\bibinfo
  {year} {2020})}\BibitemShut {NoStop}%
\bibitem [{\citenamefont {Thomas}\ \emph {et~al.}(2020)\citenamefont {Thomas},
  \citenamefont {Santos}, \citenamefont {Christensen}, \citenamefont {Asaba},
  \citenamefont {Ronning}, \citenamefont {Thompson}, \citenamefont {Bauer},
  \citenamefont {Fernandes}, \citenamefont {Fabbris},\ and\ \citenamefont
  {Rosa}}]{thomas_evidence_2020}%
  \BibitemOpen
  \bibfield  {author} {\bibinfo {author} {\bibfnamefont {S.~M.}\ \bibnamefont
  {Thomas}}, \bibinfo {author} {\bibfnamefont {F.~B.}\ \bibnamefont {Santos}},
  \bibinfo {author} {\bibfnamefont {M.~H.}\ \bibnamefont {Christensen}},
  \bibinfo {author} {\bibfnamefont {T.}~\bibnamefont {Asaba}}, \bibinfo
  {author} {\bibfnamefont {F.}~\bibnamefont {Ronning}}, \bibinfo {author}
  {\bibfnamefont {J.~D.}\ \bibnamefont {Thompson}}, \bibinfo {author}
  {\bibfnamefont {E.~D.}\ \bibnamefont {Bauer}}, \bibinfo {author}
  {\bibfnamefont {R.~M.}\ \bibnamefont {Fernandes}}, \bibinfo {author}
  {\bibfnamefont {G.}~\bibnamefont {Fabbris}}, \ and\ \bibinfo {author}
  {\bibfnamefont {P.~F.~S.}\ \bibnamefont {Rosa}},\ }\href {\doibase
  10.1126/sciadv.abc8709} {\bibfield  {journal} {\bibinfo  {journal} {Science
  Advances}\ }\textbf {\bibinfo {volume} {6}},\ \bibinfo {pages} {eabc8709}
  (\bibinfo {year} {2020})}\BibitemShut {NoStop}%
\bibitem [{\citenamefont {Aoki}\ \emph {et~al.}(2020)\citenamefont {Aoki},
  \citenamefont {Honda}, \citenamefont {Knebel}, \citenamefont {Braithwaite},
  \citenamefont {Nakamura}, \citenamefont {Li}, \citenamefont {Homma},
  \citenamefont {Shimizu}, \citenamefont {Sato}, \citenamefont {Brison},\ and\
  \citenamefont {Flouquet}}]{aoki_multiple_2021}%
  \BibitemOpen
  \bibfield  {author} {\bibinfo {author} {\bibfnamefont {D.}~\bibnamefont
  {Aoki}}, \bibinfo {author} {\bibfnamefont {F.}~\bibnamefont {Honda}},
  \bibinfo {author} {\bibfnamefont {G.}~\bibnamefont {Knebel}}, \bibinfo
  {author} {\bibfnamefont {D.}~\bibnamefont {Braithwaite}}, \bibinfo {author}
  {\bibfnamefont {A.}~\bibnamefont {Nakamura}}, \bibinfo {author}
  {\bibfnamefont {D.}~\bibnamefont {Li}}, \bibinfo {author} {\bibfnamefont
  {Y.}~\bibnamefont {Homma}}, \bibinfo {author} {\bibfnamefont
  {Y.}~\bibnamefont {Shimizu}}, \bibinfo {author} {\bibfnamefont {Y.~J.}\
  \bibnamefont {Sato}}, \bibinfo {author} {\bibfnamefont {J.-P.}\ \bibnamefont
  {Brison}}, \ and\ \bibinfo {author} {\bibfnamefont {J.}~\bibnamefont
  {Flouquet}},\ }\href {\doibase 10.7566/JPSJ.89.053705} {\bibfield  {journal}
  {\bibinfo  {journal} {Journal of the Physical Society of Japan}\ }\textbf
  {\bibinfo {volume} {89}},\ \bibinfo {pages} {053705} (\bibinfo {year}
  {2020})},\ \Eprint
  {http://arxiv.org/abs/https://doi.org/10.7566/JPSJ.89.053705}
  {https://doi.org/10.7566/JPSJ.89.053705} \BibitemShut {NoStop}%
\bibitem [{\citenamefont {Miyake}\ \emph {et~al.}(2019)\citenamefont {Miyake},
  \citenamefont {Shimizu}, \citenamefont {Sato}, \citenamefont {Li},
  \citenamefont {Nakamura}, \citenamefont {Homma}, \citenamefont {Honda},
  \citenamefont {Flouquet}, \citenamefont {Tokunaga},\ and\ \citenamefont
  {Aoki}}]{miyake_metamagnetic_2019}%
  \BibitemOpen
  \bibfield  {author} {\bibinfo {author} {\bibfnamefont {A.}~\bibnamefont
  {Miyake}}, \bibinfo {author} {\bibfnamefont {Y.}~\bibnamefont {Shimizu}},
  \bibinfo {author} {\bibfnamefont {Y.~J.}\ \bibnamefont {Sato}}, \bibinfo
  {author} {\bibfnamefont {D.}~\bibnamefont {Li}}, \bibinfo {author}
  {\bibfnamefont {A.}~\bibnamefont {Nakamura}}, \bibinfo {author}
  {\bibfnamefont {Y.}~\bibnamefont {Homma}}, \bibinfo {author} {\bibfnamefont
  {F.}~\bibnamefont {Honda}}, \bibinfo {author} {\bibfnamefont
  {J.}~\bibnamefont {Flouquet}}, \bibinfo {author} {\bibfnamefont
  {M.}~\bibnamefont {Tokunaga}}, \ and\ \bibinfo {author} {\bibfnamefont
  {D.}~\bibnamefont {Aoki}},\ }\href {\doibase 10.7566/JPSJ.88.063706}
  {\bibfield  {journal} {\bibinfo  {journal} {Journal of the Physical Society
  of Japan}\ }\textbf {\bibinfo {volume} {88}},\ \bibinfo {pages} {063706}
  (\bibinfo {year} {2019})}\BibitemShut {NoStop}%
\bibitem [{\citenamefont {Miyake}\ \emph {et~al.}(2022)\citenamefont {Miyake},
  \citenamefont {Gen}, \citenamefont {Ikeda}, \citenamefont {Miyake},
  \citenamefont {Shimizu}, \citenamefont {Sato}, \citenamefont {Li},
  \citenamefont {Nakamura}, \citenamefont {Homma}, \citenamefont {Honda},
  \citenamefont {Flouquet}, \citenamefont {Tokunaga},\ and\ \citenamefont
  {Aoki}}]{miyake_magnetovolume_2022}%
  \BibitemOpen
  \bibfield  {author} {\bibinfo {author} {\bibfnamefont {A.}~\bibnamefont
  {Miyake}}, \bibinfo {author} {\bibfnamefont {M.}~\bibnamefont {Gen}},
  \bibinfo {author} {\bibfnamefont {A.}~\bibnamefont {Ikeda}}, \bibinfo
  {author} {\bibfnamefont {K.}~\bibnamefont {Miyake}}, \bibinfo {author}
  {\bibfnamefont {Y.}~\bibnamefont {Shimizu}}, \bibinfo {author} {\bibfnamefont
  {Y.~J.}\ \bibnamefont {Sato}}, \bibinfo {author} {\bibfnamefont
  {D.}~\bibnamefont {Li}}, \bibinfo {author} {\bibfnamefont {A.}~\bibnamefont
  {Nakamura}}, \bibinfo {author} {\bibfnamefont {Y.}~\bibnamefont {Homma}},
  \bibinfo {author} {\bibfnamefont {F.}~\bibnamefont {Honda}}, \bibinfo
  {author} {\bibfnamefont {J.}~\bibnamefont {Flouquet}}, \bibinfo {author}
  {\bibfnamefont {M.}~\bibnamefont {Tokunaga}}, \ and\ \bibinfo {author}
  {\bibfnamefont {D.}~\bibnamefont {Aoki}},\ }\href {\doibase
  10.7566/JPSJ.91.063703} {\bibfield  {journal} {\bibinfo  {journal} {Journal
  of the Physical Society of Japan}\ }\textbf {\bibinfo {volume} {91}},\
  \bibinfo {pages} {063703} (\bibinfo {year} {2022})},\ \Eprint
  {http://arxiv.org/abs/https://doi.org/10.7566/JPSJ.91.063703}
  {https://doi.org/10.7566/JPSJ.91.063703} \BibitemShut {NoStop}%
\bibitem [{\citenamefont {Niu}\ \emph {et~al.}(2020)\citenamefont {Niu},
  \citenamefont {Knebel}, \citenamefont {Braithwaite}, \citenamefont {Aoki},
  \citenamefont {Lapertot}, \citenamefont {Valiska}, \citenamefont {Seyfarth},
  \citenamefont {Knafo}, \citenamefont {Helm}, \citenamefont {Brison},
  \citenamefont {Flouquet},\ and\ \citenamefont {Pourret}}]{niu_evidence_2020}%
  \BibitemOpen
  \bibfield  {author} {\bibinfo {author} {\bibfnamefont {Q.}~\bibnamefont
  {Niu}}, \bibinfo {author} {\bibfnamefont {G.}~\bibnamefont {Knebel}},
  \bibinfo {author} {\bibfnamefont {D.}~\bibnamefont {Braithwaite}}, \bibinfo
  {author} {\bibfnamefont {D.}~\bibnamefont {Aoki}}, \bibinfo {author}
  {\bibfnamefont {G.}~\bibnamefont {Lapertot}}, \bibinfo {author}
  {\bibfnamefont {M.}~\bibnamefont {Valiska}}, \bibinfo {author} {\bibfnamefont
  {G.}~\bibnamefont {Seyfarth}}, \bibinfo {author} {\bibfnamefont
  {W.}~\bibnamefont {Knafo}}, \bibinfo {author} {\bibfnamefont
  {T.}~\bibnamefont {Helm}}, \bibinfo {author} {\bibfnamefont {J.-P.}\
  \bibnamefont {Brison}}, \bibinfo {author} {\bibfnamefont {J.}~\bibnamefont
  {Flouquet}}, \ and\ \bibinfo {author} {\bibfnamefont {A.}~\bibnamefont
  {Pourret}},\ }\href {\doibase 10.1103/PhysRevResearch.2.033179} {\bibfield
  {journal} {\bibinfo  {journal} {Physical Review Research}\ }\textbf {\bibinfo
  {volume} {2}},\ \bibinfo {pages} {033179} (\bibinfo {year}
  {2020})}\BibitemShut {NoStop}%
\bibitem [{\citenamefont {Knafo}\ \emph {et~al.}(2021)\citenamefont {Knafo},
  \citenamefont {Nardone}, \citenamefont {Vališka}, \citenamefont {Zitouni},
  \citenamefont {Lapertot}, \citenamefont {Aoki}, \citenamefont {Knebel},\ and\
  \citenamefont {Braithwaite}}]{knafo_comparison_2021}%
  \BibitemOpen
  \bibfield  {author} {\bibinfo {author} {\bibfnamefont {W.}~\bibnamefont
  {Knafo}}, \bibinfo {author} {\bibfnamefont {M.}~\bibnamefont {Nardone}},
  \bibinfo {author} {\bibfnamefont {M.}~\bibnamefont {Vališka}}, \bibinfo
  {author} {\bibfnamefont {A.}~\bibnamefont {Zitouni}}, \bibinfo {author}
  {\bibfnamefont {G.}~\bibnamefont {Lapertot}}, \bibinfo {author}
  {\bibfnamefont {D.}~\bibnamefont {Aoki}}, \bibinfo {author} {\bibfnamefont
  {G.}~\bibnamefont {Knebel}}, \ and\ \bibinfo {author} {\bibfnamefont
  {D.}~\bibnamefont {Braithwaite}},\ }\href {\doibase
  10.1038/s42005-021-00545-z} {\bibfield  {journal} {\bibinfo  {journal}
  {Communications Physics}\ }\textbf {\bibinfo {volume} {4}},\ \bibinfo {pages}
  {40} (\bibinfo {year} {2021})}\BibitemShut {NoStop}%
\bibitem [{\citenamefont {Helm}\ \emph {et~al.}(2023)\citenamefont {Helm},
  \citenamefont {Kimata}, \citenamefont {Sudo}, \citenamefont {Miyata},
  \citenamefont {Stirnat}, \citenamefont {Förster}, \citenamefont {Hornung},
  \citenamefont {König}, \citenamefont {Sheikin}, \citenamefont {Pourret},
  \citenamefont {Lapertot}, \citenamefont {Aoki}, \citenamefont {Knebel},
  \citenamefont {Wosnitza},\ and\ \citenamefont {Brison}}]{helm_arxiv_2023}%
  \BibitemOpen
  \bibfield  {author} {\bibinfo {author} {\bibfnamefont {T.}~\bibnamefont
  {Helm}}, \bibinfo {author} {\bibfnamefont {M.}~\bibnamefont {Kimata}},
  \bibinfo {author} {\bibfnamefont {K.}~\bibnamefont {Sudo}}, \bibinfo {author}
  {\bibfnamefont {A.}~\bibnamefont {Miyata}}, \bibinfo {author} {\bibfnamefont
  {J.}~\bibnamefont {Stirnat}}, \bibinfo {author} {\bibfnamefont
  {T.}~\bibnamefont {Förster}}, \bibinfo {author} {\bibfnamefont
  {J.}~\bibnamefont {Hornung}}, \bibinfo {author} {\bibfnamefont
  {M.}~\bibnamefont {König}}, \bibinfo {author} {\bibfnamefont
  {I.}~\bibnamefont {Sheikin}}, \bibinfo {author} {\bibfnamefont
  {A.}~\bibnamefont {Pourret}}, \bibinfo {author} {\bibfnamefont
  {G.}~\bibnamefont {Lapertot}}, \bibinfo {author} {\bibfnamefont
  {D.}~\bibnamefont {Aoki}}, \bibinfo {author} {\bibfnamefont {G.}~\bibnamefont
  {Knebel}}, \bibinfo {author} {\bibfnamefont {J.}~\bibnamefont {Wosnitza}}, \
  and\ \bibinfo {author} {\bibfnamefont {J.-P.}\ \bibnamefont {Brison}},\
  }\href {\doibase doi.org/10.48550/arXiv.2207.08261} {\bibfield  {journal}
  {\bibinfo  {journal} {arXiv.2207.08261.}\ } (\bibinfo {year} {2023}),\
  doi.org/10.48550/arXiv.2207.08261}\BibitemShut {NoStop}%
\bibitem [{\citenamefont {Imajo}\ \emph {et~al.}(2019)\citenamefont {Imajo},
  \citenamefont {Kohama}, \citenamefont {Miyake}, \citenamefont {Dong},
  \citenamefont {Tokunaga}, \citenamefont {Flouquet}, \citenamefont {Kindo},\
  and\ \citenamefont {Aoki}}]{imajo_thermodynamic_2019}%
  \BibitemOpen
  \bibfield  {author} {\bibinfo {author} {\bibfnamefont {S.}~\bibnamefont
  {Imajo}}, \bibinfo {author} {\bibfnamefont {Y.}~\bibnamefont {Kohama}},
  \bibinfo {author} {\bibfnamefont {A.}~\bibnamefont {Miyake}}, \bibinfo
  {author} {\bibfnamefont {C.}~\bibnamefont {Dong}}, \bibinfo {author}
  {\bibfnamefont {M.}~\bibnamefont {Tokunaga}}, \bibinfo {author}
  {\bibfnamefont {J.}~\bibnamefont {Flouquet}}, \bibinfo {author}
  {\bibfnamefont {K.}~\bibnamefont {Kindo}}, \ and\ \bibinfo {author}
  {\bibfnamefont {D.}~\bibnamefont {Aoki}},\ }\href {\doibase
  10.7566/JPSJ.88.083705} {\bibfield  {journal} {\bibinfo  {journal} {Journal
  of the Physical Society of Japan}\ }\textbf {\bibinfo {volume} {88}},\
  \bibinfo {pages} {083705} (\bibinfo {year} {2019})}\BibitemShut {NoStop}%
\bibitem [{\citenamefont {Altarawneh}\ \emph {et~al.}(2009)\citenamefont
  {Altarawneh}, \citenamefont {Mielke},\ and\ \citenamefont
  {Brooks}}]{altarawneh_proximity_2009}%
  \BibitemOpen
  \bibfield  {author} {\bibinfo {author} {\bibfnamefont {M.~M.}\ \bibnamefont
  {Altarawneh}}, \bibinfo {author} {\bibfnamefont {C.~H.}\ \bibnamefont
  {Mielke}}, \ and\ \bibinfo {author} {\bibfnamefont {J.~S.}\ \bibnamefont
  {Brooks}},\ }\href {\doibase 10.1063/1.3152219} {\bibfield  {journal}
  {\bibinfo  {journal} {Review of Scientific Instruments}\ }\textbf {\bibinfo
  {volume} {80}},\ \bibinfo {pages} {066104} (\bibinfo {year}
  {2009})}\BibitemShut {NoStop}%
\bibitem [{\citenamefont {Ghannadzadeh}\ \emph {et~al.}(2011)\citenamefont
  {Ghannadzadeh}, \citenamefont {Coak}, \citenamefont {Franke}, \citenamefont
  {Goddard}, \citenamefont {Singleton},\ and\ \citenamefont
  {Manson}}]{ghannadzadeh_measurement_2011}%
  \BibitemOpen
  \bibfield  {author} {\bibinfo {author} {\bibfnamefont {S.}~\bibnamefont
  {Ghannadzadeh}}, \bibinfo {author} {\bibfnamefont {M.}~\bibnamefont {Coak}},
  \bibinfo {author} {\bibfnamefont {I.}~\bibnamefont {Franke}}, \bibinfo
  {author} {\bibfnamefont {P.~A.}\ \bibnamefont {Goddard}}, \bibinfo {author}
  {\bibfnamefont {J.}~\bibnamefont {Singleton}}, \ and\ \bibinfo {author}
  {\bibfnamefont {J.~L.}\ \bibnamefont {Manson}},\ }\href {\doibase
  10.1063/1.3653395} {\bibfield  {journal} {\bibinfo  {journal} {Review of
  Scientific Instruments}\ }\textbf {\bibinfo {volume} {82}},\ \bibinfo {pages}
  {113902} (\bibinfo {year} {2011})}\BibitemShut {NoStop}%
\bibitem [{\citenamefont {Jaime}\ \emph {et~al.}(2002)\citenamefont {Jaime},
  \citenamefont {Kim}, \citenamefont {Jorge}, \citenamefont {McCall},\ and\
  \citenamefont {Mydosh}}]{jaime_high_2002}%
  \BibitemOpen
  \bibfield  {author} {\bibinfo {author} {\bibfnamefont {M.}~\bibnamefont
  {Jaime}}, \bibinfo {author} {\bibfnamefont {K.~H.}\ \bibnamefont {Kim}},
  \bibinfo {author} {\bibfnamefont {G.}~\bibnamefont {Jorge}}, \bibinfo
  {author} {\bibfnamefont {S.}~\bibnamefont {McCall}}, \ and\ \bibinfo {author}
  {\bibfnamefont {J.~A.}\ \bibnamefont {Mydosh}},\ }\href {\doibase
  10.1103/PhysRevLett.89.287201} {\bibfield  {journal} {\bibinfo  {journal}
  {Physical Review Letters}\ }\textbf {\bibinfo {volume} {89}},\ \bibinfo
  {pages} {287201} (\bibinfo {year} {2002})}\BibitemShut {NoStop}%
\bibitem [{\citenamefont {Silhanek}\ \emph {et~al.}(2006)\citenamefont
  {Silhanek}, \citenamefont {Jaime}, \citenamefont {Harrison}, \citenamefont
  {Fanelli}, \citenamefont {Batista}, \citenamefont {Amitsuka}, \citenamefont
  {Nakatsuji}, \citenamefont {Balicas}, \citenamefont {Kim}, \citenamefont
  {Fisk}, \citenamefont {Sarrao}, \citenamefont {Civale},\ and\ \citenamefont
  {Mydosh}}]{silhanek_irreversible_2006}%
  \BibitemOpen
  \bibfield  {author} {\bibinfo {author} {\bibfnamefont {A.~V.}\ \bibnamefont
  {Silhanek}}, \bibinfo {author} {\bibfnamefont {M.}~\bibnamefont {Jaime}},
  \bibinfo {author} {\bibfnamefont {N.}~\bibnamefont {Harrison}}, \bibinfo
  {author} {\bibfnamefont {V.~R.}\ \bibnamefont {Fanelli}}, \bibinfo {author}
  {\bibfnamefont {C.~D.}\ \bibnamefont {Batista}}, \bibinfo {author}
  {\bibfnamefont {H.}~\bibnamefont {Amitsuka}}, \bibinfo {author}
  {\bibfnamefont {S.}~\bibnamefont {Nakatsuji}}, \bibinfo {author}
  {\bibfnamefont {L.}~\bibnamefont {Balicas}}, \bibinfo {author} {\bibfnamefont
  {K.~H.}\ \bibnamefont {Kim}}, \bibinfo {author} {\bibfnamefont
  {Z.}~\bibnamefont {Fisk}}, \bibinfo {author} {\bibfnamefont {J.~L.}\
  \bibnamefont {Sarrao}}, \bibinfo {author} {\bibfnamefont {L.}~\bibnamefont
  {Civale}}, \ and\ \bibinfo {author} {\bibfnamefont {J.~A.}\ \bibnamefont
  {Mydosh}},\ }\href {\doibase 10.1103/PhysRevLett.96.136403} {\bibfield
  {journal} {\bibinfo  {journal} {Physical Review Letters}\ }\textbf {\bibinfo
  {volume} {96}},\ \bibinfo {pages} {136403} (\bibinfo {year}
  {2006})}\BibitemShut {NoStop}%
\bibitem [{\citenamefont {Smylie}\ \emph {et~al.}(2019)\citenamefont {Smylie},
  \citenamefont {Koshelev}, \citenamefont {Willa}, \citenamefont {Willa},
  \citenamefont {Kwok}, \citenamefont {Bao}, \citenamefont {Chung},
  \citenamefont {Kanatzidis}, \citenamefont {Singleton}, \citenamefont
  {Balakirev}, \citenamefont {Hebbeker}, \citenamefont {Niraula}, \citenamefont
  {Bokari}, \citenamefont {Kayani},\ and\ \citenamefont {Welp}}]{SMYLIE}%
  \BibitemOpen
  \bibfield  {author} {\bibinfo {author} {\bibfnamefont {M.~P.}\ \bibnamefont
  {Smylie}}, \bibinfo {author} {\bibfnamefont {A.~E.}\ \bibnamefont
  {Koshelev}}, \bibinfo {author} {\bibfnamefont {K.}~\bibnamefont {Willa}},
  \bibinfo {author} {\bibfnamefont {R.}~\bibnamefont {Willa}}, \bibinfo
  {author} {\bibfnamefont {W.-K.}\ \bibnamefont {Kwok}}, \bibinfo {author}
  {\bibfnamefont {J.-K.}\ \bibnamefont {Bao}}, \bibinfo {author} {\bibfnamefont
  {D.~Y.}\ \bibnamefont {Chung}}, \bibinfo {author} {\bibfnamefont {M.~G.}\
  \bibnamefont {Kanatzidis}}, \bibinfo {author} {\bibfnamefont
  {J.}~\bibnamefont {Singleton}}, \bibinfo {author} {\bibfnamefont {F.~F.}\
  \bibnamefont {Balakirev}}, \bibinfo {author} {\bibfnamefont {H.}~\bibnamefont
  {Hebbeker}}, \bibinfo {author} {\bibfnamefont {P.}~\bibnamefont {Niraula}},
  \bibinfo {author} {\bibfnamefont {E.}~\bibnamefont {Bokari}}, \bibinfo
  {author} {\bibfnamefont {A.}~\bibnamefont {Kayani}}, \ and\ \bibinfo {author}
  {\bibfnamefont {U.}~\bibnamefont {Welp}},\ }\href {\doibase
  10.1103/PhysRevB.100.054507} {\bibfield  {journal} {\bibinfo  {journal}
  {Phys. Rev. B}\ }\textbf {\bibinfo {volume} {100}},\ \bibinfo {pages}
  {054507} (\bibinfo {year} {2019})}\BibitemShut {NoStop}%
\bibitem [{Note1()}]{Note1}%
  \BibitemOpen
  \bibinfo {note} {In the PDO circuit, the resonant frequency $f = 1/\protect
  \sqrt {LC}$, where $L$ is the circuit inductance and $C$ the capacitance. The
  inductance of a solenoid of $N$ turns, area $A$, and length $l$ is $L=\mu
  N^2A/l$, where $\mu $ is the permeability of the core material. Due to the
  skin-depth effect in metallic samples of low resistance the effective volume,
  and hence inductance, of the coil is reduced. Hence, $\Delta f \propto
  (\Delta \rho )^{-1}$. In UTe$_2$ this frequency shift is most evident upon
  entering the superconducting state where the penetration depth is
  significantly smaller than the normal state skin depth.}\BibitemShut {Stop}%
\bibitem [{\citenamefont {Jaime}\ \emph {et~al.}(2000)\citenamefont {Jaime},
  \citenamefont {Movshovich}, \citenamefont {Steward}, \citenamefont
  {Beyermann}, \citenamefont {Gomez~Berisso}, \citenamefont {Hundley},
  \citenamefont {Canfield},\ and\ \citenamefont {Sarrao}}]{jaime_closing_2000}%
  \BibitemOpen
  \bibfield  {author} {\bibinfo {author} {\bibfnamefont {M.}~\bibnamefont
  {Jaime}}, \bibinfo {author} {\bibfnamefont {R.}~\bibnamefont {Movshovich}},
  \bibinfo {author} {\bibfnamefont {G.}~\bibnamefont {Steward}}, \bibinfo
  {author} {\bibfnamefont {W.}~\bibnamefont {Beyermann}}, \bibinfo {author}
  {\bibfnamefont {M.}~\bibnamefont {Gomez~Berisso}}, \bibinfo {author}
  {\bibfnamefont {M.}~\bibnamefont {Hundley}}, \bibinfo {author} {\bibfnamefont
  {P.}~\bibnamefont {Canfield}}, \ and\ \bibinfo {author} {\bibfnamefont
  {J.}~\bibnamefont {Sarrao}},\ }\href {\doibase 10.1038/35012027} {\bibfield
  {journal} {\bibinfo  {journal} {Nature}\ }\textbf {\bibinfo {volume} {405}},\
  \bibinfo {pages} {160} (\bibinfo {year} {2000})}\BibitemShut {NoStop}%
\bibitem [{\citenamefont {Jaime}(2010)}]{jaime_frontiers_2009}%
  \BibitemOpen
  \bibfield  {author} {\bibinfo {author} {\bibfnamefont {M.}~\bibnamefont
  {Jaime}},\ }\href@noop {} {\bibfield  {journal} {\bibinfo  {journal} {Netsu
  Sokutei}\ }\textbf {\bibinfo {volume} {37}},\ \bibinfo {pages} {26} (\bibinfo
  {year} {2010})}\BibitemShut {NoStop}%
\bibitem [{\citenamefont {Ravex}\ \emph {et~al.}(1987)\citenamefont {Ravex},
  \citenamefont {Flouquet}, \citenamefont {Tholence}, \citenamefont {Jaccard},\
  and\ \citenamefont {Meyer}}]{ravex_1987}%
  \BibitemOpen
  \bibfield  {author} {\bibinfo {author} {\bibfnamefont {A.}~\bibnamefont
  {Ravex}}, \bibinfo {author} {\bibfnamefont {J.}~\bibnamefont {Flouquet}},
  \bibinfo {author} {\bibfnamefont {J.}~\bibnamefont {Tholence}}, \bibinfo
  {author} {\bibfnamefont {D.}~\bibnamefont {Jaccard}}, \ and\ \bibinfo
  {author} {\bibfnamefont {A.}~\bibnamefont {Meyer}},\ }\href {\doibase
  10.1016/0304-8853(87)90621-4} {\bibfield  {journal} {\bibinfo  {journal}
  {Journal of Magnetism and Magnetic Materials}\ }\textbf {\bibinfo {volume}
  {63-64}},\ \bibinfo {pages} {400} (\bibinfo {year} {1987})}\BibitemShut
  {NoStop}%
\bibitem [{\citenamefont {Suderow}\ \emph {et~al.}(1997)\citenamefont
  {Suderow}, \citenamefont {Brison}, \citenamefont {Huxley},\ and\
  \citenamefont {J.}}]{suderow_1997}%
  \BibitemOpen
  \bibfield  {author} {\bibinfo {author} {\bibfnamefont {H.}~\bibnamefont
  {Suderow}}, \bibinfo {author} {\bibfnamefont {J.}~\bibnamefont {Brison}},
  \bibinfo {author} {\bibfnamefont {A.}~\bibnamefont {Huxley}}, \ and\ \bibinfo
  {author} {\bibfnamefont {F.}~\bibnamefont {J.}},\ }\href
  {https://link.springer.com/article/10.1007/BF02396814#Bib1} {\bibfield
  {journal} {\bibinfo  {journal} {Journal of Low Temperature Physics}\ }\textbf
  {\bibinfo {volume} {108}},\ \bibinfo {pages} {11} (\bibinfo {year}
  {1997})}\BibitemShut {NoStop}%
\bibitem [{\citenamefont {Howald}(2006)}]{howald_PhD}%
  \BibitemOpen
  \bibfield  {author} {\bibinfo {author} {\bibfnamefont {L.}~\bibnamefont
  {Howald}},\ }\emph {\bibinfo {title} {Interactions entre la
  supraconductivit\'{e} et la criticit\'{e} quantique, dans les composes
  CeCoIn$_5$, URhGe et UCoGe}},\ \href@noop {} {Ph.D. thesis},\ \bibinfo
  {school} {Universit\'{e} de Grenoble}, \bibinfo {address} {Grenoble, France}
  (\bibinfo {year} {2006})\BibitemShut {NoStop}%
\bibitem [{\citenamefont {Taupin}\ \emph {et~al.}(2014)\citenamefont {Taupin},
  \citenamefont {Howald}, \citenamefont {Aoki},\ and\ \citenamefont
  {Brison}}]{taupin_2014}%
  \BibitemOpen
  \bibfield  {author} {\bibinfo {author} {\bibfnamefont {M.}~\bibnamefont
  {Taupin}}, \bibinfo {author} {\bibfnamefont {L.}~\bibnamefont {Howald}},
  \bibinfo {author} {\bibfnamefont {D.}~\bibnamefont {Aoki}}, \ and\ \bibinfo
  {author} {\bibfnamefont {J.-P.}\ \bibnamefont {Brison}},\ }\href {\doibase
  10.1103/PhysRevB.90.180501} {\bibfield  {journal} {\bibinfo  {journal} {Phys.
  Rev. B}\ }\textbf {\bibinfo {volume} {90}},\ \bibinfo {pages} {180501}
  (\bibinfo {year} {2014})}\BibitemShut {NoStop}%
\bibitem [{\citenamefont {Nikolo}\ \emph {et~al.}(2018)\citenamefont {Nikolo},
  \citenamefont {Singleton}, \citenamefont {Solenov}, \citenamefont {Jiang},
  \citenamefont {Weiss},\ and\ \citenamefont {Hellstrom}}]{NIKOLO}%
  \BibitemOpen
  \bibfield  {author} {\bibinfo {author} {\bibfnamefont {M.}~\bibnamefont
  {Nikolo}}, \bibinfo {author} {\bibfnamefont {J.}~\bibnamefont {Singleton}},
  \bibinfo {author} {\bibfnamefont {D.}~\bibnamefont {Solenov}}, \bibinfo
  {author} {\bibfnamefont {J.}~\bibnamefont {Jiang}}, \bibinfo {author}
  {\bibfnamefont {J.}~\bibnamefont {Weiss}}, \ and\ \bibinfo {author}
  {\bibfnamefont {E.}~\bibnamefont {Hellstrom}},\ }\href {\doibase
  https://doi.org/10.1016/j.physb.2017.09.047} {\bibfield  {journal} {\bibinfo
  {journal} {Physica B: Condensed Matter}\ }\textbf {\bibinfo {volume} {536}},\
  \bibinfo {pages} {833} (\bibinfo {year} {2018})}\BibitemShut {NoStop}%
\bibitem [{\citenamefont {Singleton}\ \emph {et~al.}(2010)\citenamefont
  {Singleton}, \citenamefont {de~la Cruz}, \citenamefont {McDonald},
  \citenamefont {Li}, \citenamefont {Altarawneh}, \citenamefont {Goddard},
  \citenamefont {Franke}, \citenamefont {Rickel}, \citenamefont {Mielke},
  \citenamefont {Yao},\ and\ \citenamefont {Dai}}]{CdC}%
  \BibitemOpen
  \bibfield  {author} {\bibinfo {author} {\bibfnamefont {J.}~\bibnamefont
  {Singleton}}, \bibinfo {author} {\bibfnamefont {C.}~\bibnamefont {de~la
  Cruz}}, \bibinfo {author} {\bibfnamefont {R.~D.}\ \bibnamefont {McDonald}},
  \bibinfo {author} {\bibfnamefont {S.}~\bibnamefont {Li}}, \bibinfo {author}
  {\bibfnamefont {M.}~\bibnamefont {Altarawneh}}, \bibinfo {author}
  {\bibfnamefont {P.}~\bibnamefont {Goddard}}, \bibinfo {author} {\bibfnamefont
  {I.}~\bibnamefont {Franke}}, \bibinfo {author} {\bibfnamefont
  {D.}~\bibnamefont {Rickel}}, \bibinfo {author} {\bibfnamefont {C.~H.}\
  \bibnamefont {Mielke}}, \bibinfo {author} {\bibfnamefont {X.}~\bibnamefont
  {Yao}}, \ and\ \bibinfo {author} {\bibfnamefont {P.}~\bibnamefont {Dai}},\
  }\href {\doibase 10.1103/PhysRevLett.104.086403} {\bibfield  {journal}
  {\bibinfo  {journal} {Phys. Rev. Lett.}\ }\textbf {\bibinfo {volume} {104}},\
  \bibinfo {pages} {086403} (\bibinfo {year} {2010})}\BibitemShut {NoStop}%
\bibitem [{\citenamefont {Xiang}\ \emph {et~al.}(2021)\citenamefont {Xiang},
  \citenamefont {Chen}, \citenamefont {Chen}, \citenamefont {Tinsman},
  \citenamefont {Sato}, \citenamefont {Asaba}, \citenamefont {Lu},
  \citenamefont {Kasahara}, \citenamefont {Jaime}, \citenamefont {Balakirev},
  \citenamefont {Iga}, \citenamefont {Matsuda}, \citenamefont {Singleton},\
  and\ \citenamefont {Li}}]{Xiang2021}%
  \BibitemOpen
  \bibfield  {author} {\bibinfo {author} {\bibfnamefont {Z.}~\bibnamefont
  {Xiang}}, \bibinfo {author} {\bibfnamefont {L.}~\bibnamefont {Chen}},
  \bibinfo {author} {\bibfnamefont {K.-W.}\ \bibnamefont {Chen}}, \bibinfo
  {author} {\bibfnamefont {C.}~\bibnamefont {Tinsman}}, \bibinfo {author}
  {\bibfnamefont {Y.}~\bibnamefont {Sato}}, \bibinfo {author} {\bibfnamefont
  {T.}~\bibnamefont {Asaba}}, \bibinfo {author} {\bibfnamefont
  {H.}~\bibnamefont {Lu}}, \bibinfo {author} {\bibfnamefont {Y.}~\bibnamefont
  {Kasahara}}, \bibinfo {author} {\bibfnamefont {M.}~\bibnamefont {Jaime}},
  \bibinfo {author} {\bibfnamefont {F.}~\bibnamefont {Balakirev}}, \bibinfo
  {author} {\bibfnamefont {F.}~\bibnamefont {Iga}}, \bibinfo {author}
  {\bibfnamefont {Y.}~\bibnamefont {Matsuda}}, \bibinfo {author} {\bibfnamefont
  {J.}~\bibnamefont {Singleton}}, \ and\ \bibinfo {author} {\bibfnamefont
  {L.}~\bibnamefont {Li}},\ }\href {\doibase 10.1038/s41567-021-01216-0}
  {\bibfield  {journal} {\bibinfo  {journal} {Nature Physics}\ }\textbf
  {\bibinfo {volume} {17}},\ \bibinfo {pages} {788} (\bibinfo {year}
  {2021})}\BibitemShut {NoStop}%
\bibitem [{\citenamefont {G\"otze}\ \emph {et~al.}(2020)\citenamefont
  {G\"otze}, \citenamefont {Pearce}, \citenamefont {Goddard}, \citenamefont
  {Jaime}, \citenamefont {Maple}, \citenamefont {Sasmal}, \citenamefont
  {Yanagisawa}, \citenamefont {McCollam}, \citenamefont {Khouri}, \citenamefont
  {Ho},\ and\ \citenamefont {Singleton}}]{PCHo}%
  \BibitemOpen
  \bibfield  {author} {\bibinfo {author} {\bibfnamefont {K.}~\bibnamefont
  {G\"otze}}, \bibinfo {author} {\bibfnamefont {M.~J.}\ \bibnamefont {Pearce}},
  \bibinfo {author} {\bibfnamefont {P.~A.}\ \bibnamefont {Goddard}}, \bibinfo
  {author} {\bibfnamefont {M.}~\bibnamefont {Jaime}}, \bibinfo {author}
  {\bibfnamefont {M.~B.}\ \bibnamefont {Maple}}, \bibinfo {author}
  {\bibfnamefont {K.}~\bibnamefont {Sasmal}}, \bibinfo {author} {\bibfnamefont
  {T.}~\bibnamefont {Yanagisawa}}, \bibinfo {author} {\bibfnamefont
  {A.}~\bibnamefont {McCollam}}, \bibinfo {author} {\bibfnamefont
  {T.}~\bibnamefont {Khouri}}, \bibinfo {author} {\bibfnamefont {P.-C.}\
  \bibnamefont {Ho}}, \ and\ \bibinfo {author} {\bibfnamefont {J.}~\bibnamefont
  {Singleton}},\ }\href {\doibase 10.1103/PhysRevB.101.075102} {\bibfield
  {journal} {\bibinfo  {journal} {Phys. Rev. B}\ }\textbf {\bibinfo {volume}
  {101}},\ \bibinfo {pages} {075102} (\bibinfo {year} {2020})}\BibitemShut
  {NoStop}%
\bibitem [{\citenamefont {Stokes}\ \emph {et~al.}(1984)\citenamefont {Stokes},
  \citenamefont {Bloch}, \citenamefont {Janossy},\ and\ \citenamefont
  {Gruner}}]{stokes_1984}%
  \BibitemOpen
  \bibfield  {author} {\bibinfo {author} {\bibfnamefont {J.}~\bibnamefont
  {Stokes}}, \bibinfo {author} {\bibfnamefont {A.}~\bibnamefont {Bloch}},
  \bibinfo {author} {\bibfnamefont {A.}~\bibnamefont {Janossy}}, \ and\
  \bibinfo {author} {\bibfnamefont {G.}~\bibnamefont {Gruner}},\ }\href
  {https://journals.aps.org/prl/pdf/10.1103/PhysRevLett.52.372} {\bibfield
  {journal} {\bibinfo  {journal} {Phys. Rev. Lett.}\ }\textbf {\bibinfo
  {volume} {52}},\ \bibinfo {pages} {372} (\bibinfo {year} {1984})}\BibitemShut
  {NoStop}%
\bibitem [{Note2()}]{Note2}%
  \BibitemOpen
  \bibinfo {note} {We note a discrepancy in the entropy change at the
  metamagnetic transition between the present results and the values reported
  in the brief report by Imajo \protect \textit {et al.} Ref.~\cite
  {imajo_thermodynamic_2019}. The difference is likely related to the
  conditions in the present work, \protect \textit {i.e.} a stronger link to
  the thermal bath needed to reach lower $^3$He temperatures leading to the
  quasi-adiabatic magnetization of the sample. The difference does not affect
  the conclusions of the current paper.}\BibitemShut {Stop}%
\bibitem [{\citenamefont {Miyake}\ \emph {et~al.}(2021)\citenamefont {Miyake},
  \citenamefont {Shimizu}, \citenamefont {Sato}, \citenamefont {Li},
  \citenamefont {Nakamura}, \citenamefont {Homma}, \citenamefont {Honda},
  \citenamefont {Flouquet}, \citenamefont {Tokunaga},\ and\ \citenamefont
  {Aoki}}]{miyake_enhancement_2021}%
  \BibitemOpen
  \bibfield  {author} {\bibinfo {author} {\bibfnamefont {A.}~\bibnamefont
  {Miyake}}, \bibinfo {author} {\bibfnamefont {Y.}~\bibnamefont {Shimizu}},
  \bibinfo {author} {\bibfnamefont {Y.~J.}\ \bibnamefont {Sato}}, \bibinfo
  {author} {\bibfnamefont {D.}~\bibnamefont {Li}}, \bibinfo {author}
  {\bibfnamefont {A.}~\bibnamefont {Nakamura}}, \bibinfo {author}
  {\bibfnamefont {Y.}~\bibnamefont {Homma}}, \bibinfo {author} {\bibfnamefont
  {F.}~\bibnamefont {Honda}}, \bibinfo {author} {\bibfnamefont
  {J.}~\bibnamefont {Flouquet}}, \bibinfo {author} {\bibfnamefont
  {M.}~\bibnamefont {Tokunaga}}, \ and\ \bibinfo {author} {\bibfnamefont
  {D.}~\bibnamefont {Aoki}},\ }\href {\doibase 10.7566/JPSJ.90.103702}
  {\bibfield  {journal} {\bibinfo  {journal} {Journal of the Physical Society
  of Japan}\ }\textbf {\bibinfo {volume} {90}},\ \bibinfo {pages} {103702}
  (\bibinfo {year} {2021})}\BibitemShut {NoStop}%
\bibitem [{\citenamefont {Wu}\ \emph {et~al.}(2023)\citenamefont {Wu},
  \citenamefont {Weinberger}, \citenamefont {Chen}, \citenamefont {Cabala},
  \citenamefont {Chichinadze}, \citenamefont {Shaffer}, \citenamefont
  {Pospisil}, \citenamefont {Prokleska}, \citenamefont {Haidamak},
  \citenamefont {Bastien}, \citenamefont {Sechovsky}, \citenamefont {Hickey},
  \citenamefont {Mancera-Ugarte}, \citenamefont {Benjamin}, \citenamefont
  {Graf}, \citenamefont {Skourski}, \citenamefont {Lonzarich}, \citenamefont
  {Valiska}, \citenamefont {Grosche},\ and\ \citenamefont
  {Eaton}}]{wu_arxiv_2023}%
  \BibitemOpen
  \bibfield  {author} {\bibinfo {author} {\bibfnamefont {Z.}~\bibnamefont
  {Wu}}, \bibinfo {author} {\bibfnamefont {T.}~\bibnamefont {Weinberger}},
  \bibinfo {author} {\bibfnamefont {J.}~\bibnamefont {Chen}}, \bibinfo {author}
  {\bibfnamefont {A.}~\bibnamefont {Cabala}}, \bibinfo {author} {\bibfnamefont
  {D.}~\bibnamefont {Chichinadze}}, \bibinfo {author} {\bibfnamefont
  {D.}~\bibnamefont {Shaffer}}, \bibinfo {author} {\bibfnamefont
  {J.}~\bibnamefont {Pospisil}}, \bibinfo {author} {\bibfnamefont
  {J.}~\bibnamefont {Prokleska}}, \bibinfo {author} {\bibfnamefont
  {T.}~\bibnamefont {Haidamak}}, \bibinfo {author} {\bibfnamefont
  {G.}~\bibnamefont {Bastien}}, \bibinfo {author} {\bibfnamefont
  {V.}~\bibnamefont {Sechovsky}}, \bibinfo {author} {\bibfnamefont
  {A.}~\bibnamefont {Hickey}}, \bibinfo {author} {\bibfnamefont
  {M.}~\bibnamefont {Mancera-Ugarte}}, \bibinfo {author} {\bibfnamefont
  {S.}~\bibnamefont {Benjamin}}, \bibinfo {author} {\bibfnamefont {D.~E.}\
  \bibnamefont {Graf}}, \bibinfo {author} {\bibfnamefont {Y.}~\bibnamefont
  {Skourski}}, \bibinfo {author} {\bibfnamefont {G.}~\bibnamefont {Lonzarich}},
  \bibinfo {author} {\bibfnamefont {M.}~\bibnamefont {Valiska}}, \bibinfo
  {author} {\bibfnamefont {F.}~\bibnamefont {Grosche}}, \ and\ \bibinfo
  {author} {\bibfnamefont {A.}~\bibnamefont {Eaton}},\ }\href {\doibase
  doi.org/10.48550/arXiv.2305.19033} {\bibfield  {journal} {\bibinfo  {journal}
  {arXiv.2305.19033}\ } (\bibinfo {year} {2023}),\
  doi.org/10.48550/arXiv.2305.19033}\BibitemShut {NoStop}%
\bibitem [{\citenamefont {Frank}\ \emph {et~al.}(2023)\citenamefont {Frank},
  \citenamefont {Lewin}, \citenamefont {Saucedo~Salas}, \citenamefont {Czajka},
  \citenamefont {Hayes}, \citenamefont {Yoon}, \citenamefont {Metz},
  \citenamefont {Paglione}, \citenamefont {Singleton},\ and\ \citenamefont
  {Butch}}]{frank_arxiv_2023}%
  \BibitemOpen
  \bibfield  {author} {\bibinfo {author} {\bibfnamefont {C.}~\bibnamefont
  {Frank}}, \bibinfo {author} {\bibfnamefont {S.}~\bibnamefont {Lewin}},
  \bibinfo {author} {\bibfnamefont {G.}~\bibnamefont {Saucedo~Salas}}, \bibinfo
  {author} {\bibfnamefont {P.}~\bibnamefont {Czajka}}, \bibinfo {author}
  {\bibfnamefont {I.}~\bibnamefont {Hayes}}, \bibinfo {author} {\bibfnamefont
  {H.}~\bibnamefont {Yoon}}, \bibinfo {author} {\bibfnamefont {T.}~\bibnamefont
  {Metz}}, \bibinfo {author} {\bibfnamefont {J.}~\bibnamefont {Paglione}},
  \bibinfo {author} {\bibfnamefont {J.}~\bibnamefont {Singleton}}, \ and\
  \bibinfo {author} {\bibfnamefont {N.}~\bibnamefont {Butch}},\ }\href
  {\doibase doi.org/10.48550/arXiv.2304.12392} {\bibfield  {journal} {\bibinfo
  {journal} {arXiv.2304.12392}\ } (\bibinfo {year} {2023}),\
  doi.org/10.48550/arXiv.2304.12392}\BibitemShut {NoStop}%
\bibitem [{\citenamefont {Rosa}\ \emph {et~al.}(2022)\citenamefont {Rosa},
  \citenamefont {Weiland}, \citenamefont {Fender}, \citenamefont {Scott},
  \citenamefont {Ronning}, \citenamefont {Thompson}, \citenamefont {Bauer},\
  and\ \citenamefont {Thomas}}]{rosa_single-component_2021}%
  \BibitemOpen
  \bibfield  {author} {\bibinfo {author} {\bibfnamefont {P.~F.~S.}\
  \bibnamefont {Rosa}}, \bibinfo {author} {\bibfnamefont {A.}~\bibnamefont
  {Weiland}}, \bibinfo {author} {\bibfnamefont {S.~S.}\ \bibnamefont {Fender}},
  \bibinfo {author} {\bibfnamefont {B.~L.}\ \bibnamefont {Scott}}, \bibinfo
  {author} {\bibfnamefont {F.}~\bibnamefont {Ronning}}, \bibinfo {author}
  {\bibfnamefont {J.~D.}\ \bibnamefont {Thompson}}, \bibinfo {author}
  {\bibfnamefont {E.~D.}\ \bibnamefont {Bauer}}, \ and\ \bibinfo {author}
  {\bibfnamefont {S.~M.}\ \bibnamefont {Thomas}},\ }\href {\doibase
  10.1038/s43246-022-00254-2} {\bibfield  {journal} {\bibinfo  {journal}
  {Communications Materials}\ }\textbf {\bibinfo {volume} {3}},\ \bibinfo
  {pages} {33} (\bibinfo {year} {2022})}\BibitemShut {NoStop}%
\bibitem [{\citenamefont {Willis}\ \emph {et~al.}(2020)\citenamefont {Willis},
  \citenamefont {Ding}, \citenamefont {Singleton},\ and\ \citenamefont
  {Balakirev}}]{goniometer}%
  \BibitemOpen
  \bibfield  {author} {\bibinfo {author} {\bibfnamefont {X.}~\bibnamefont
  {Willis}}, \bibinfo {author} {\bibfnamefont {X.}~\bibnamefont {Ding}},
  \bibinfo {author} {\bibfnamefont {J.}~\bibnamefont {Singleton}}, \ and\
  \bibinfo {author} {\bibfnamefont {F.~F.}\ \bibnamefont {Balakirev}},\ }\href
  {\doibase 10.1063/1.5125792} {\bibfield  {journal} {\bibinfo  {journal}
  {Review of Scientific Instruments}\ }\textbf {\bibinfo {volume} {91}},\
  \bibinfo {pages} {036102} (\bibinfo {year} {2020})}\BibitemShut {NoStop}%
\end{thebibliography}%

\end{document}